# Frequency support Scheme based on parametrized power curve for de-loaded Wind Turbine under various wind speed

Cheng Zhong[12], Yueming Lv[1], Huayi Li[1], JiKai Chen[1], Yang Li[1]
[1] Key Laboratory of Modern Power System Simulation and Control & Renewable Energy Technology （Ministry of Education）, Northeast Electric Power University, Jilin ,132012, China；
[2] Department of Electronic, Electrical and Systems Engineering, University of Birmingham, Birmingham B15 2TT, U.K
Corresponding author: Cheng Zhong
Email: zhongcheng@neepu.edu.cn

**Abstract: With increased wind power penetration in modern power systems, wind plants are required to provide frequency support similar to conventional plants. However, for the existing frequency regulation scheme of wind turbines, the control gains in the auxiliary frequency controller are difficult to set because of the compromise of the frequency regulation performance and the stable operation of wind turbines, especially when the wind speed remains variable. This paper proposes a novel frequency regulation scheme (FRS) for de-loaded wind turbines. Instead of an auxiliary frequency controller, frequency support is provided by modifying the parametrized power versus rotor speed (Pw-ωr) curve, including the inertia power versus rotor speed curve and the droop power versus rotor speed curve. The advantage of the proposed scheme is that it does not contain any control gains and generally adapts to different wind speeds. Further, the proposed scheme can work for the whole section of wind speed without wind speed measurement information. The compared simulation results demonstrate the scheme improves the system frequency response while ensuring the stable operation of doubly-fed induction generators (DFIGs)-based variable-speed wind turbines (VSWTs) under various wind conditions. Furthermore, the scheme prevents rotor speed overdeceleration even when the wind speed decreases during frequency regulation control.**

## 1. Introduction

Wind power generation is the most popular renewable generation technology, and the technology of wind turbine is still improving [1-3], such as the improvement of wind turbine cooling system [4-6], fault analysis [7-8], and so on. In 2020, the new installation of wind power generation was 93 Gw, and the total installed capacity was 743 Gw [9]. Approximately 95% of installed wind turbines (WTs) are Variable speed wind turbines (VSWTs), either DFIGs-based with partially rated converters or PMSGs-based with fully rated converters [10,11]. Unlike conventional power generators, VSWTs have no inherent inertial response because of the power electrical converter interface. VSWTs usually do not participate in the system frequency response for operation in maximum power point tracking (MPPT) mode. Therefore, as the penetration of VSWTs increases, the inertial and frequency regulation ability of the whole power system will degrade, causing frequency stability issues [12,13]. Some countries have required wind plants to provide frequency support [14-16].

Many research studies have discussed the frequency regulation scheme (FRS) for VSWTs. The strategies can be classified into inertial response control and de-loaded control [17,18].

For inertial response control, the VSWTs still operate in MPPT mode, and the rotational kinetic energy (KE) of VSWTs is released to deliver temporary addition power during frequency dips. Further, the inertial response control can divide in two subcategories [17]: natural inertial control [19-26] and stepwise control [27-29].

For natural inertial control, the value of the addition power is determined by the frequency measurement, such as the rate of change of frequency (ROCOF) [19], or the frequency deviation [20] [21], or both of them [22-26]. Considering that the wind speed is variable and the change of the rotor speed is complicated, the auxiliary frequency controller's gains should be selected carefully with the trade-off considering the frequency regulation performance and the stable operating range of the wind turbine. Therefore, some varying gain methods have been suggested. In [22], the control gains of FRS under different wind speeds is adjusted based on the wind speed. However, the pre-determined gains are obtained by the off-line modeling analysis, and the wind speed measurement may not be obtained or inaccuracy. To improve the frequency nadir (FN) and ensure stable operation of DFIG, the droop gains is dynamically changes based on ROCOF in [23]. In [24], the gains of additional ROCOF and frequency deviation loops is adaptively tuned depended on the rotor speed measurement. [25] present a time varying gains determined based on desired frequency -response time to raise frequency nadir and eliminate frequency second dip. [26] proposed an adaptive droop gain which is a function of real-time rotor speed and wind power penetration level.

For step-wise control, the addition frequency power is determined by the pre-set power surge function, such as step function [27], ramp function [28] or torque limit function [29]. Compared with natural inertial response, the inertial power using the step-wise control can be properly tuned according to different shapes in terms of its magnitude and duration. An optimization approaches employing the genetic algorithm are proposed to maximize the released energy from the wind turbine during its overproduction period [30].However, during the rotor speed recovery period, the output power of wind turbine reduced and may cause a secondary frequency drop [31, 32].In [31,32],the incremental power varies with the rotor speed and wind power penetration levels during the overproduction period, and then, the reference power smoothly decreases with time and rotor speed during recovery period.

For inertial response control, because of the limit of the rotor kinetic energy, it only affords for seconds-term frequency support. While, for the de-loaded control, VSWTs reserve a part of the active power through pitch angle control [33,34], over-speed control [35], or combination of both [36-38]. It can provide a minutes-term primary frequency support.

Due to rotor speed limit, the over-speed de-loaded control is only adapted for low wind range. In [39,40], three wind speed modes are defined: low wind speed mode where de-loaded operation is merely by rotor speed control; medium wind speed mode where de-loaded operation is conducted by combining pitch angle control and rotor speed control; and high wind speed mode where modified pitch angle control alone. However, it required accuracy wind speed information to judge the wind speed mode, and the calculation of de-loaded power reference need both parameters of wind turbine and wind speed. [41] present a variable droop control strategy that considers optional rotor kinetic energy. However, the rotor kinetic energy estimation required the wind speed information and parameters of wind turbine. [42] proposed a comprehensive frequency control that combines the temporary power injection control and power reserve control with consider rotor security and maximum extricable energy of wind turbines. But, it still required the parameter of wind turbines. In [43] a comprehensive frequency regulation that combines the step-wise inertial control and variable-droop control is present.

Actually, in most of the literatures mentioned above, the additional frequency regulation power is determined by an auxiliary ROCOF and frequency deviation loops, which is added to the de-loaded power reference. The gains of the auxiliary frequency controller are difficult to set a proper value compromising of the frequency regulation performance and wind turbines rotor security. Moreover, if these schemes are applied to multiple WTGs, difficulties will arise in determining the different gains for all WTGs. Nevertheless, some adaptive gains methods in [22-26,41,42], the proper initial value or parameters are also difficult to select [22-26], or the gains are determined by evaluating available energy which of the accurate parameters of wind turbines and wind speed information are required [41,42].

This paper proposes a novel frequency support scheme for DFIG-based wind turbines. Instead of the auxiliary frequency controller in the most existing scheme, the additional frequency regulation power for wind turbines is determined by the modified parametrized power versus rotor speed curve. There are three main advantages of the proposed control scheme:

(1) There are no control gains in the scheme. Thus, it does not need to carefully select a proper control gains for wind turbines like the existing scheme. It can generally adapt for multiple WTGs with different wind speeds.

(2) It has potential self-adaptive frequency support with wind speed and can continuously ensure that the wind turbine operates within the safe rotor speed range, even in the case of a sudden decrease in wind speed.

(3) It can work for the whole wind speed range and does not need wind speed information.

The remainder of this paper is organized as follows: In Section 2, the DFIG-based wind turbine model and the traditional frequency regulation scheme are introduced. In Section 3, the proposed frequency regulation scheme for DFIG-based wind turbines is presented. In Section 4, compared with the traditional frequency regulation, the proposed control scheme's performance is demonstrated under various wind conditions. Finally, a brief conclusion is drawn in Section 5.

## 2. DFIG Model of a DFIG-based Wind generation

Fig. 1 shows the block diagram of DFIG-based wind turbines' simplified model, commonly used for frequency control studies and developed in [45]. The mechanical power of a DFIG captured from the wind, $P_{m,}$ is defined by [46]

$$P_m = \frac{1}{2}\rho\pi R^2 V_W{}^3 C_p\left(\lambda,\beta\right) \quad (1)$$

where $\rho$—air density, $R$—radius, $V_w$—wind speed, λ—tip speed ratio, $\lambda=\omega_r R/V_w$, $\omega_r$—rotor speed, $\beta$—pitch angle, and $Cp(\lambda,\beta)$—power coefficient.

As in [47], $C_p(\lambda,\beta)$ can be expressed as:

$$C_p\left(\lambda,\beta\right) = (0.44-0.0167\beta)\sin[\frac{\pi(\lambda-3)}{5-0.3\beta}]-0.00184(\lambda-3)\beta \quad (2)$$

A "one mass" system has been considered to represent the rotational dynamics of the gearbox, wind turbine and electrical generator [48, 49], whose equivalent moment of inertia is $J_{eq.}$ where $T$, $P$, and $\omega$ represent torque, power and angular speed, respectively; subscripts $g$ and $t$ are used to indicate the variables referring to the generator and the turbine. $fs$, $p$ and $n$ are the grid frequency, the number of pole pairs and the gear ratio of the DFIG, respectively.

In addition, the DFIG and the rotor side converter (RSC) are both regarded as a single first-order dynamics actuator, with a time constant $\tau_C$, whose input is the electromagnetic reference torque from the speed control system $T_g{}^*$.

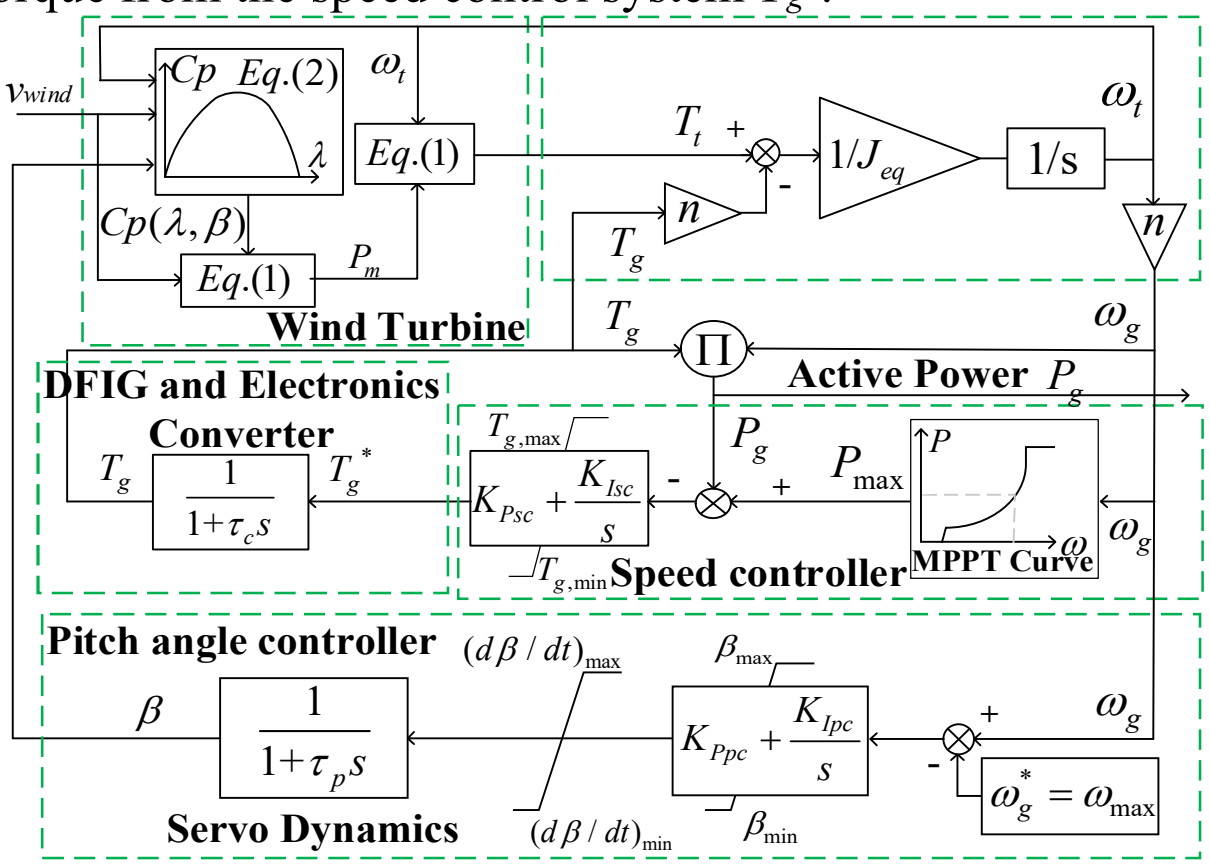

***Fig. 1** Block diagram of a simplified model of a DFIG-based wind turbine*

### 2.1. Maximum Power Point Tracking Controller (MPPT)

To capture the maximum wind power by the wind turbine, a power reference, $P_{max}$ is from the maximum power versus rotor speed ($P_{max}$-$\omega_r$) curve that can be represented by (3) and illustrated in Fig. 2 (the solid black line).

$$
P_{\max}\text{-}\omega_r = \begin{cases} \dfrac{k_{opt}\omega_1^{\ 3}}{\omega_1-\omega_{\min}}\left(\omega_r-\omega_{\min}\right) & \omega_{\min}\le\omega_r<\omega_0 \\ k_{opt}\omega_r^{\ 3} & \omega_0\le\omega_r<\omega_1 \\ \dfrac{P_{nor}-k_{opt}\omega_2^{\ 3}}{\left(\omega_{\max}-\omega_2\right)}\left(\omega_r-\omega_{\max}\right)+P_{nor} & \omega_1\le\omega_r<\omega_{\max} \\ P_{nor} & \omega_{\max}\le\omega_r \end{cases} \tag{3}
$$

where $k_{opt}$ is the optimization constant, whose value depends on the physical characteristics of the wind turbine.

Concerning Fig. 2, the maximum power curve is divided into four segments according to the rotor speed. The segment A-B corresponds to the starting zone. In segment B-D, known as the optimization zone, the rotor speed is adjusted to the optimal speed with the optimal power coefficient $Cp(\lambda,\beta)$. Segments D-E are constant rotor speed zones, and the rotor is almost invariable. After the segment after point E is called the constant power zone, $P_{max}$ is constant $P_{nor}$.

The intersection point of the capture power versus rotor speed curve ($P_m$-$\omega_r$) and the maximum power curve ($P_{max}$-$\omega_r$) is an equilibrium point. After some disturbances, the DFIG-WT automatically converges to the intersection point, where the captured mechanical power $P_m$ is equal to the optimum power $P_{opt}$, and the rotor equals $\omega_{opt}$.

A pitch-angle controller is used to prevent the rotor speed from exceeding $\omega_{max}$ and keep the output power at the rated value when wind speeds are high.

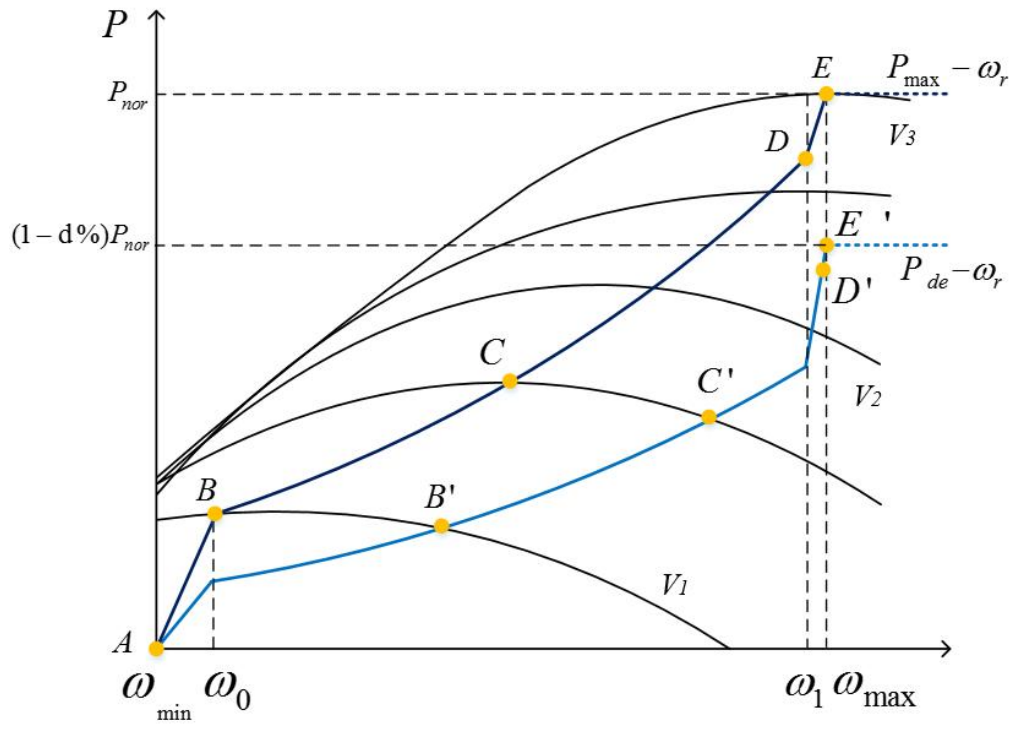

***Fig. 2** Maximum power curve and de-loaded power curve for DFIG-based VSWTs*

## 2.2. The traditional FRS for DFIG-based WT

The most popular FRS for wind turbines is shown in Fig. 3, as in [50] (and similar schemes in [22-26], [36-38]). To realize de-loaded control, a de-loaded power versus rotor speed curve ($P_{de}$-$\omega_r$) replaces the $P_{max}$-$\omega_r$ curve. This makes the rotor speed higher than the optimum rotor speed. The DFIG-WT operates at a suboptimal point below the maximum power point (reserve a part of the active power). A typical $P_{de}$-$\omega_r$ curve is given in Fig. 2 (the blue line).

An auxiliary frequency controller (AFC) is added to generate an additional power $\Delta P_f$, as expressed in (4). It includes virtual inertia response and droop response. The inertia response is based on the ROCOF, while the droop response is based on the frequency deviation.

$$
\Delta P_f = K_v \cdot \frac{d\Delta f_s}{dt} + \frac{1}{R}\cdot \Delta f_s \tag{4}
$$

where $K_v$ and $1/R$ are the gains of the virtual inertia and droop loops, respectively.

The pitch angle control not only prevents the rotor speed from exceeding $\omega_{max}$ but also helps to realize de-loaded operation at medium and high wind speeds area [50]. According to wind speed, the de-loaded control can be divided into three areas in [40], as shown in Fig. 2.

(1) Low wind speed area: $V_1$-$V_2$, $\beta_{de}$=0, $\omega_r$<$\omega_{max}$; only the speed control loop is used to realize de-loaded control.

(2) Medium wind speed area: $V_2$-$V_3$, $\beta_{de}$>0, $\omega_r$<$\omega_{max}$. Both the speed control loop and pitch controller are used to realize de-loaded control.

(3) High wind area: higher than $V_3$, $\beta_{de}$>0, $\omega_r$=$\omega_{max}$, only the pitch controller is used to realize de-loaded control. $\beta_{de}$ is the de-loaded pitch reference for avoiding overspeed.

The control of the pitch angle is shown in Fig. 3(b). $\beta_{de}$ is obtained from wind turbine modeling by solving Eq. (5).

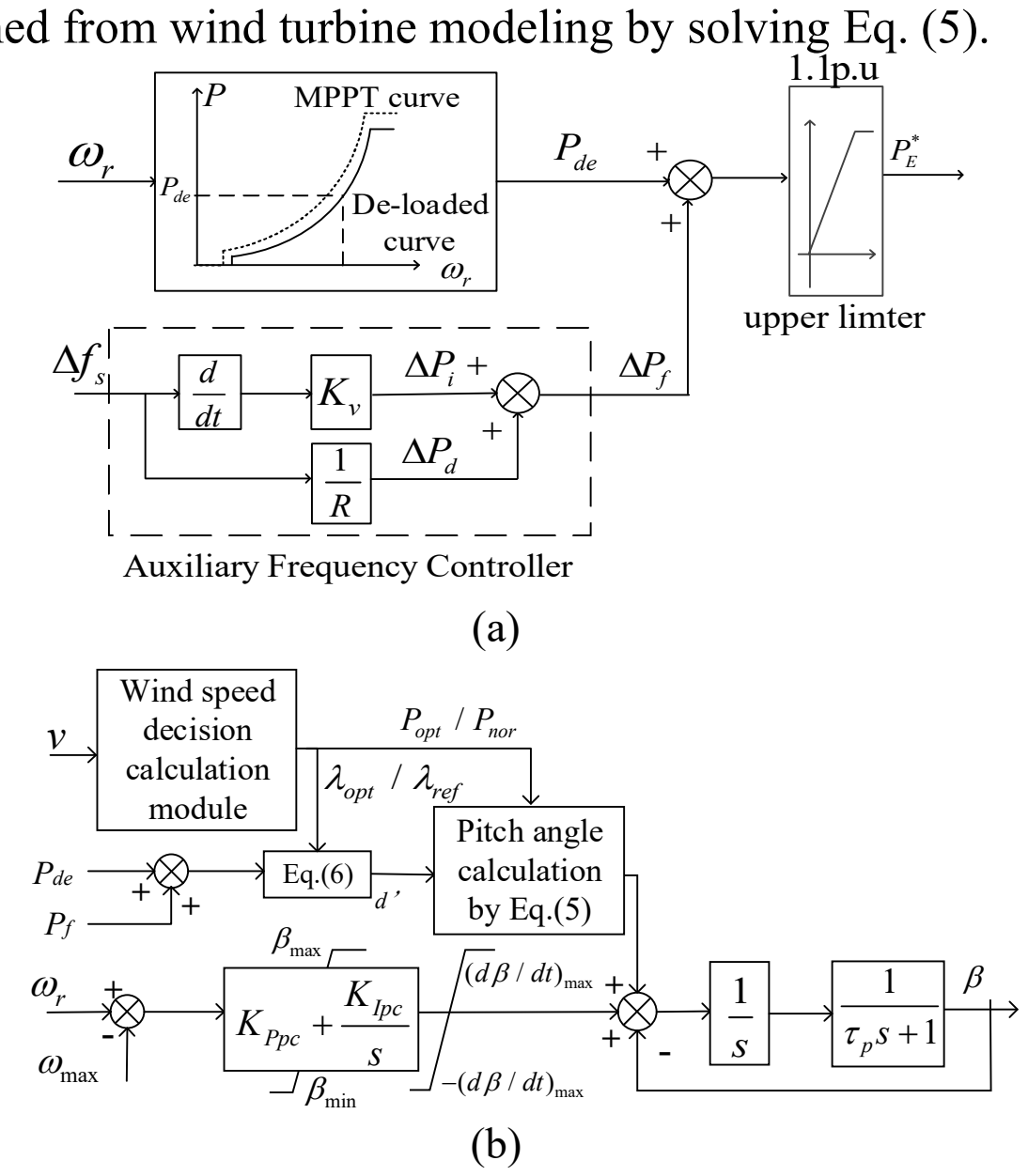

***Fig. 3** The commonly used frequency regulation controller for DFIG-based wind turbine*

$$
\begin{cases} \beta_{ref}=0 & v_1<v<v_2 \\ C_{p,de}\left(\lambda_{ref},\beta_{\text{de}}\right)=\left(1-d'\%\right)C_{p,\max}\left(\lambda_{opt},0\right) & v_2<v<v_3 \\ C_{p,de}\left(\lambda_{ref},\beta_{de}\right)=\left(1-d'\%\right)C_{p,\text{rated}}\left(\lambda_{ref},\beta_0\right) & v_3<v \end{cases} \tag{5}
$$

where $\lambda_{opt}$ is the optimum tip speed ratio and $C_{p,max}$ is the maximum wind energy capture factor. $\lambda_{ref}$ is the reference tip speed ratio in over speed control, $\lambda_{ref}=\omega_{max}R/v$, and $C_{p,rated}$ is the wind energy capture coefficient when operating at rated power. $\beta_0$ is the pitch angle reference when the wind turbines operate at the rated power. $\beta_{de}$ is the de-loaded pitch angle in the de-loaded control mode. $d'\%$ is the real de-loaded ratio of the wind turbine and can be calculated as:

$$
d' = \begin{cases} 1-\dfrac{P_{de}+\Delta P_f}{P_{opt}}, v\le V_3 \\ 1-\dfrac{P_{de}+\Delta P_f}{P_{nor}}, v> V_3 \end{cases} \tag{6}
$$

where $P_{opt}$ is the reference power of the wind turbine under MPPT and $P_{nor}$ is the rated power.

However, there are still some shortcomings for the traditional frequency regulation scheme.

(1) The gains of AFC are challenging to set because of the compromise of the frequency support performance and wind turbines' stable operation. A large gain can improve the frequency regulation while causing overdeceleration of a DFIG. Conversely, a small gain can prevent overdeceleration, but it provides a limited contribution to frequency supports. In addition, for multiple WTGs, the available energy is different because the available energy is determined by the wind turbine characteristics and wind speed. There cannot be a single proper value for all different DFIGs.

(2) Wind speed information is required to realize de-loaded and frequency support control for the whole section of wind speed. As seen in Fig. 3(b), wind speed information $V$ is necessary for the decision of the wind speed area, calculating the maximum power $P_{opt}$ and the optimum tip speed ratio $\lambda_{opt}$.

## 3. The Proposed Frequency Regulation Scheme

To address the limitations mentioned above, a novel frequency regulation scheme for de-loaded wind turbines is proposed. The whole control diagram of the proposed scheme is illustrated in Fig. 4.

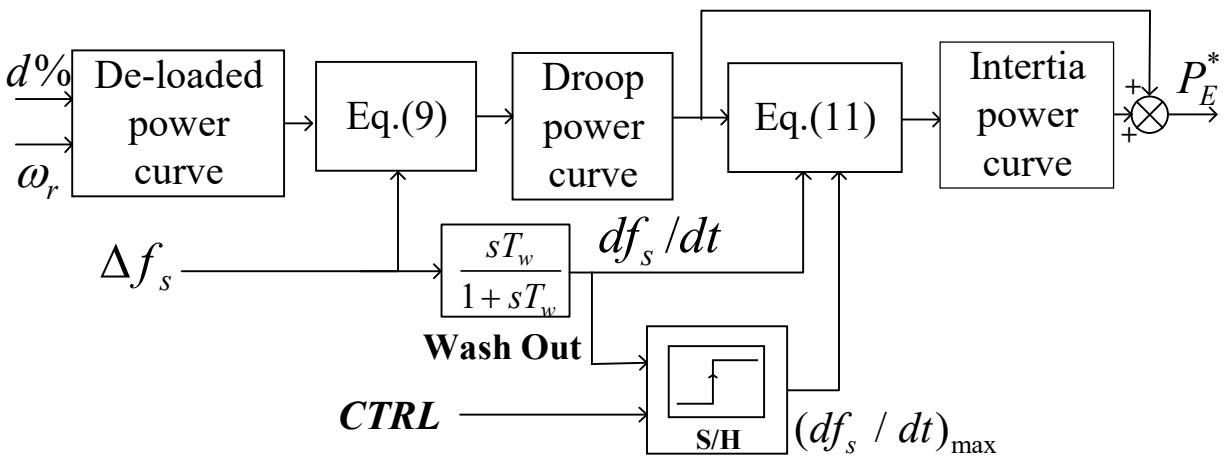

***Fig. 4*** *The proposed frequency regulation scheme*

The proposed FRS includes two key steps. First, the de-loaded curve is modified into a droop power curve based on the frequency deviation to provide a droop frequency response. Second, the droop power curve is further modified into an inertia power curve based on ROCOF to provide both inertia and droop frequency responses. The detection of d$f_s$/dt is sensitive to noise and harmonic disturbance. Hence, a washout filter ($T_w$=0.01) [51] is used to obtain d$f_s$/dt, as seen in Fig.4.

Then, the power reference $P_E$* can be obtained based on the measurement of the rotor speed $\omega_r$.

The de-loaded power curve, the droop power curve and the inertia power curve are detailed below.

### 3.1. De-loaded Power Curve

The fixed power reserve d%, is defined as

$$\mathrm{d}\% = \frac{P_{opt} - P_{de}}{P_{opt}} \qquad (7)$$

where $P_{opt}$ is the maximum available power, $P_{de}$ is the de-loaded power, and $d\%$ is set to 10% in the paper.

Similar to the $P_{max}$-$\omega_r$ curve, the de-loaded curve $P_{de}$-$\omega_r$ curve can be similarly expressed：

$$P_{de} - \omega_r = \begin{cases} \dfrac{k_{de}\omega_1^3}{\omega_1 - \omega_{\min}}(\omega_r - \omega_{\min}) & \omega_{\min} \le \omega_r < \omega_0 \\ k_{de}\omega_r^3 & \omega_0 \le \omega_r < \omega_1 \\ \dfrac{0.9P_{nor} - k_{de}\omega_2^3}{(\omega_{\max} - \omega_2)}(\omega_r - \omega_{\max}) + 0.9P_{nor} & \omega_1 \le \omega_r < \omega_{\max} \\ 0.9P_{nor} & \omega_{\max} \le \omega_r \end{cases} \qquad (8)$$

where $k_{de}$ is the de-loaded constant and $P_{de}$ is the de-loaded power.

Noticed that $k_{de}$ does not equal the 0.9$k_{opt}$. As shown in Fig. 4, for the same wind speed, the rotor speed with 0.9$P_{opt}$ is larger than the optimum rotor speed. By off-line data fitting, it can be obtained that the value of $k_{de}$ with a 10% power reserved ratio is 0.2172. At the maximum rotor speed for the de-loaded power curve, the corresponding wind speed is 10 m/s, instead of 12 m/s for the maximum power curve.

### 3.2. Droop Power versus rotor Curve

The droop power curve, $P_{droop}$-$\omega_r$, is shifted from the de-loaded curve ($P_{de}$-$\omega_r$) to the maximum power curve ($P_{max}$-$\omega_r$) based on the frequency deviation.

*d%* is 10% in this paper, and the droop curve $P_{droop}$-$\omega_r$ is defined as follows:

$$P_{droop} - \omega_r = \begin{cases} \dfrac{k_{sopt}\omega_1^3}{\omega_1 - \omega_{\min}}(\omega_r - \omega_{\min}) & \omega_{\min} \le \omega_r < \omega_0 \\ k_{sopt}\omega_r^3 & \omega_0 \le \omega_r < \omega_1 \\ \dfrac{\left(\left(0.9 - 0.1\dfrac{\Delta f}{\Delta f_{\max}}\right)\cdot P_{nor} - k_{sopt}\omega_1^3\right)}{(\omega_{\max} - \omega_1)}(\omega_r - \omega_{\max}) + P_{nor} & \omega_1 \le \omega_r < \omega_{\max} \\ \left(0.9 - 0.1\dfrac{\Delta f}{\Delta f_{\max}}\right)P_{nor} & \omega_{\max} \le \omega_r \end{cases} \qquad (9)$$

where $k_{sopt}$ is defined as,

$$k_{sopt} = \begin{cases} k_{de} - \left(k_{opt} - k_{de}\right)\dfrac{\Delta f}{\Delta f_{\max}} & \Delta f < 0 \\ k_{de} - \left(k_{de} - k_{de}^{80\%}\right)\dfrac{\Delta f}{\Delta f_{\max}} & \Delta f > 0 \end{cases} \qquad (10)$$

Where $k_{de}^{80\%}$ is the de-loaded constant with an 80% power reserve for the wind turbine, which to provide 10% power regulation capability for frequency rise event. $k_{de}^{80\%}$ is obtains by off-line data fitting and $k_{de}^{80\%}$=0.1956. $\Delta f_{max}$ is the allowable frequency deviation and $\Delta f_{max}$=0.5Hz in this paper.

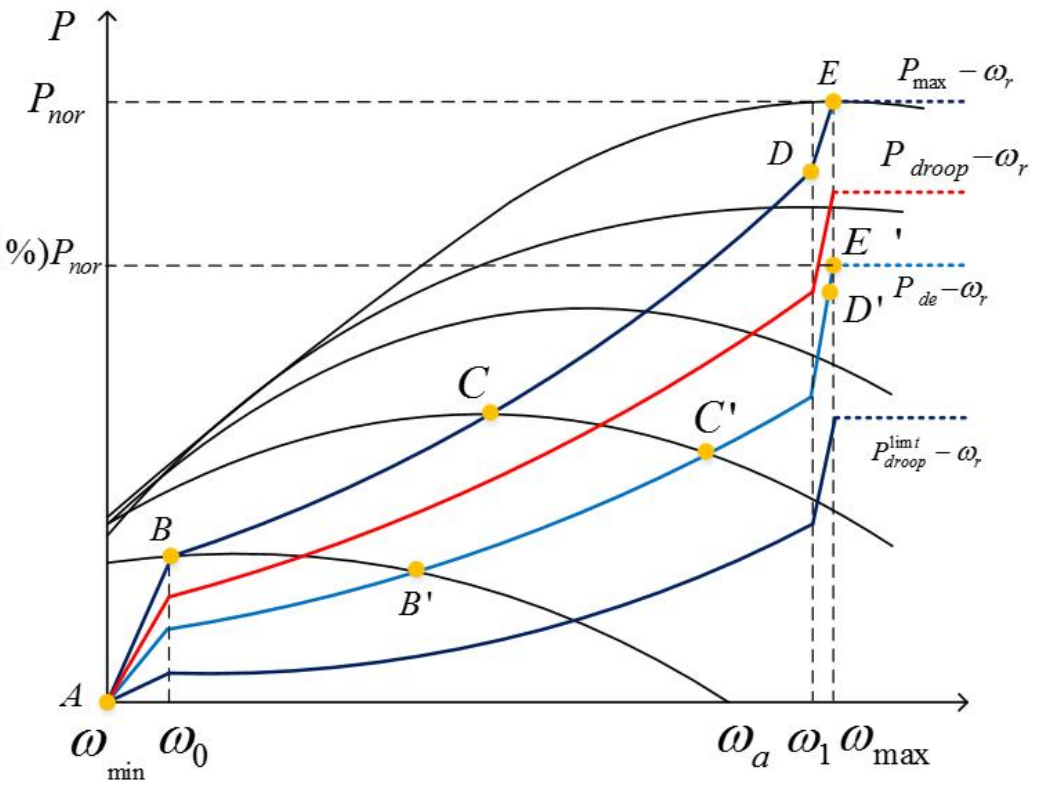

***Fig. 5*** *The droop power curve*

The droop power curve is illustrated in Fig. 5.

When the frequency dips, the droop curve moves toward the $P_{max}$-$\omega_r$ curve. Thus, the reference power with the same rotor speed is larger, i.e., more active power from the wind turbine is delivered to the system.

The droop response is a minute-term time response. Thus, the $P_{max}$-$\omega_r$ curve is the upper limit of the droop power curve. The additional power from the $P_{droop}$-$\omega_r$ curve does not exceed the maximum available power.

When $\Delta f_s$=0, $P_{droop}$-$\omega_r$ is the same as the de-loaded power curve. Otherwise, when the frequency rises, the power curve moves down, and the reference power with the same rotor speed changes to a smaller value. $P_{droop}^{limt}$ -$\omega_r$ is the lower limit droop curve (as seen in Fig. 5) and is near 80% of the maximum power curve.

Therefore, the droop power curve provides droop frequency support for the wind turbine, similar to the droop response in the traditional auxiliary frequency controller.

Note that there is always an intersection of the droop power curve and the captured wind power curve (equilibrium point). This means that the DFIG always converges to the equilibrium point in any case. Furthermore, the droop power curve definition does not contain any control gains in (9).

### 3.3. Inertia Power Curve

Furthermore, to provide inertia supports for the system frequency, the droop power curve, $P_{droop}$-$\omega_r$, is further modified into an inertia power versus rotor speed curve, called $P_{in}$-$\omega_r$, based on the ROCOF ($df_s/dt$).

In this paper, the inertia power curve is defined in (11) as below.

$$P_{in}-\omega_r=\begin{cases}\left(P_{up}^{inertial}(\omega_r)-P_{droop}(\omega_r)\right)\left|\dfrac{(df_s/dt)}{(df_s/dt)_{\max}}\right| & df_s/dt<0,\Delta f<0\\ \left(P_{lower}^{inertial}(\omega_r)-P_{droop}(\omega_r)\right)\left|\dfrac{(df_s/dt)}{(df_s/dt)_{\max}}\right| & df_s/dt>0,\Delta f>0\\ P_{droop}(\omega_r) & others\end{cases} \tag{11}$$

where $P_{up}^{inertial}$-$\omega_r$ is the upper limit of the inertia power curve, and $P_{lower}^{inertial}$-$\omega_r$ is the lower limit of the inertia power curve. $(df_s/dt)_{\max}$ is the maximum measurement $df_s/dt$ during the frequency event process.

To obtain $(df_s/dt)_{\max}$, a latch is used to store the maximum value. This means that if a new measurement ($df_s$/dt) value is larger than the old storage value, the new value replaces the old storage value. Otherwise, the latch keeps the old storage value.

$P_{up}^{inertial}$-$\omega_r$ is defined as in (12), which borrows from [42]. $P_{up}^{inertial}$-$\omega_r$ also shown in Fig. 6.

$$P_{up}^{inertial}(\omega_r)=\frac{P_{up}^{T\lim}(\omega_a)-P_{de}(\omega_0)}{\omega_a-\omega_0}(\omega_r-\omega_0)+P_{de}(\omega_0) \tag{12}$$

$P_{Tlim}$ is the torque limit relative to the power curve. $\omega_a$ is the initial rotor speed

To avoid the rapid and excessive increase in the output power causing the wind turbine's mechanical torsion, the power limit $P_{limit}$ and the maximum torque limit $T_g^{max}$ are often set to 1.1 pu. and 1.07pu [32].

The minimum torque limit $T_g^{min}$ are often set to 0. 05pu [45]. The rotor speed increases during the frequency increase event. Therefore, the definition of the $P_{lower}^{inertial}$-$\omega_r$ is only considered the range of the rotor speed higher than the current rotor $\omega_a$ . Similarly, the definition of $P_{lower}^{inertial}$-$\omega_r$ is required to prevent the speed rotor over-accelerating. the $P_{up}^{inertial}$-$\omega_r$ is given in (13), and as the red curve shown in Fig. 6.

$$P_{lower}^{inertial}(\omega_r)=\frac{P_{lower}^{T\lim}(\omega_a)-k_{de}^{80\%}\omega_1^3}{\omega_a-\omega_1}(\omega_r-\omega_1)+k_{de}^{80\%}\omega_1^3 \tag{13}$$

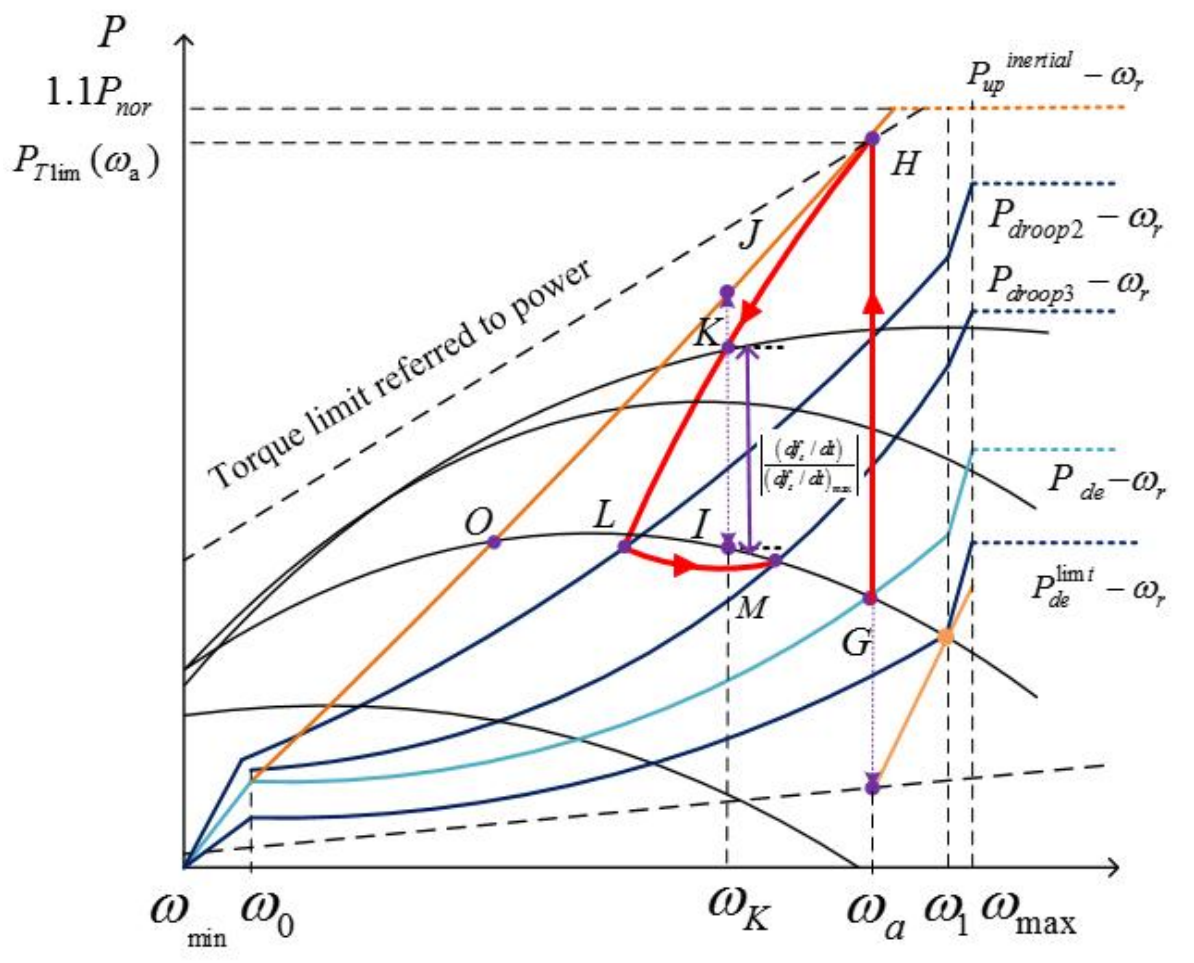

***Fig. 6*** *The inertia power curve*

To explain the frequency regulation proceeding, a frequency drop event is taken as an example, and the trajectory of operating point during the frequency regulation process is depicted by the red curve in Fig.6.

At the initial time, 'G' is assumed to be the initial operating point located at the de-loaded curve. $\Delta f_s$ and $|df_s/dt|$ are negative values, and $df_s$/dt quickly drops to the minimum value. That is, $|df_s/dt|$ reaches the maximum value. The de-loaded power curve quickly turns into the upper limit inertial curve, $P_{up}^{inertial}$-$\omega_r$. Thus, the operating point switch from the 'G' to the point 'H'. The wind turbine releases the maximum allowable power for the supported frequency.

Then, $\Delta f_s$ and $|df_s/dt|$ decreases. According to (9), the droop power curve moves up with $\Delta f_s$. Meanwhile, the rotor speed $\omega_r$ decreases because of the extra active power releasing. 'J' is the corresponding point with $\omega_K$ located at $P_{up}^{inertial}$-$\omega_r$ curve, while 'I' is the corresponding point with $\omega_K$ located at $P_{droop}$-$\omega_r$ curve. Because of the decrease of the $df_s$/dt ($df_s/dt<(df_s/dt)_{max}$), according to (11), the operating point moved from 'H' to 'K' ($P_{in}(\omega_K)$,$\omega_K$). The additional active power from the wind turbine is gradually decreased. Along with $df_s$/dt approach to zero, the operating point moves from the $P_{up}^{inertial}$-$\omega_r$ curve to the droop power curve.

When $df_s$/dt=0, the frequency reaches the lowest point. At this time, the inertia power curve turns into the droop power curve. The operating point 'K' will turn into 'L' located into the droop power curve $P_{droop2}$-$\omega_r$.

Then, the frequency recovery starts. $df_s$/dt and $\Delta f_s$ are opposite in sign. Unfortunately, the inertia response is not beneficial for frequency recovery. Thus, when $df_s$/dt >0 and $\Delta f_s$ <0, the power curve maintains the droop power curve. The inertial power curve is only enabled when $df_s$ <0 and $df_s$/dt<0. As the frequency recovery ($\Delta f_s$ increase), the $P_{droop2}$-$\omega_r$ curve will move downward to the $P_{droop3}$-$\omega_r$ curve. As illustrated in Fig.6, the operating point will move from 'L' to 'M'.

The $P_{up}^{inertial}$-$\omega_r$ curve and the $P_m$-$\omega r$ curve always has an intersection (like 'O' point in Fig.6), which also is an equilibrium point. Only under the extreme situations that d$f_s$/dt keeps the max value (d$f_s$/dt)$_{max}$, the wind turbine will converge to this intersection. The rotor speed of this intersection still higher the minimum rotor speed limit $\omega_{min}$. Practically, the d$f_s$/dt will gradually reduce during the frequency regulation process. Hence, the rotor speed always higher than $\omega_{min.}$ during frequency regulation process. Otherwise, the proceeding of the frequency rise event is similar. The proposed method can ensure the wind turbine operate among the safe rotor range and prevents overdeceleration.

Noticeably, there are no control gains in (9) - (11). The output power is determined by the current measurement rotor speed and the modified power curve. Thus, unlike the existing scheme's difficulty in choosing the proper control gains, the proposed scheme does not have any control gains.

### 3.4. Pitch Angle Control

As described in the traditional FRS control (Section II. B), in the medium and high wind areas, pitch angle control is necessary to add to limit the rotor speed and help de-load control. However, in traditional control, wind speed information is required to decide the wind speed area and calculate the value of the compensation pitch $\beta_{de}$. However, the inaccurate wind speed measurement may be harmful to the control performance. Furthermore, a complex calculation is needed to calculate $\beta_{de}$.In this paper, improved pitch angle control is designed.

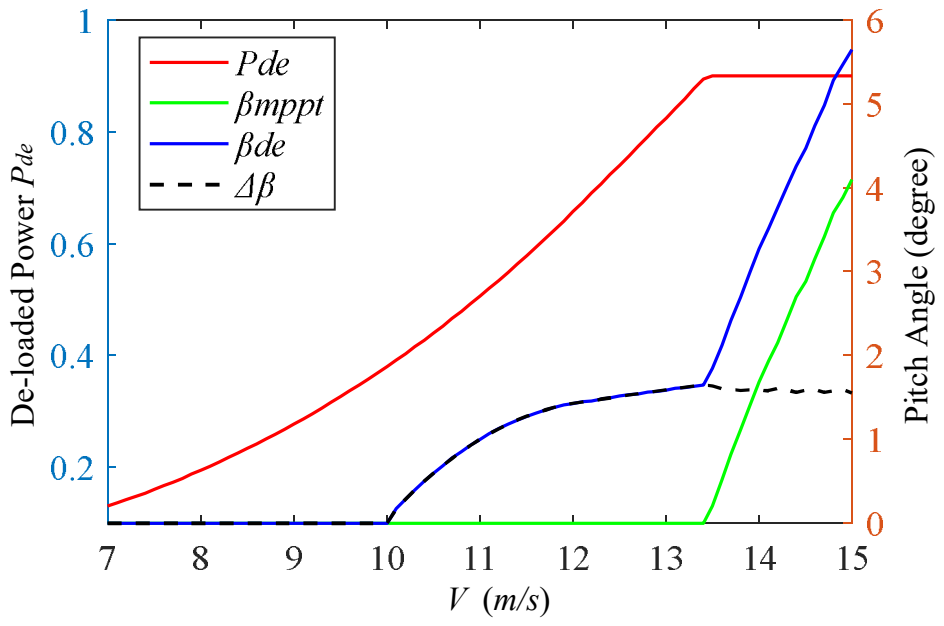

***Fig.** 7 Pitch angle and de-loaded power at various wind speeds*

Fig. 7 shows the de-loaded power, maximum power pitch angle $\beta_m$, de-loaded pitch angle $\beta_{de}$, and difference angle (between the aforementioned two angles) $\Delta\beta$ versus wind speed. The maximum power pitch angle remains zero until the wind speed reaches $V_2$. Then, the pitch angle gradually increases to reduce the capture of wind power.

The de-loaded pitch angle remains zero until the wind speed reaches $V_1$. Then, the pitch angle increases to realize de-loaded control. During $V_1$ and $V_2$, which is called the medium wind speed area aforementioned, both pitch angle control and rotor speed control are used to realize de-loaded control. At wind speeds higher than $V_2$, only the pitch angle is used to realize de-loaded control.

Observing $\Delta\beta$ in Fig. 7, $\Delta\beta$ can be divided into three segments: a low wind speed area where the wind speed lowers $V_1$, where it is zero; a medium wind speed area during $V_1$ and $V_2$, where it is a nonlinear curve; and a high wind speed area, higher than $V_2$, where it has a constant value (nearly 1.6° in this paper).

Fig. 8 shows the wind power coefficient $C_p$ versus pitch angle under different tip speed ratios.

In the vicinity of $\lambda_{opt}$ ($\lambda_{opt}$=10.5 in this paper), $C_p$ seems not to be influenced by the different $\lambda$. Therefore, the pitch angle is an almost constant value when $C_p$ is not a considerable reduction. This is the reason that in the high wind area, $\Delta\beta$ remains almost constant to realize a certain power reserve.

Linearly fitting the $\lambda_{opt}$ curve to obtain the approximate linear relationship between $C_p$ and pitch angle:

$$C_p = -0.0276\beta + 0.44 \tag{14}$$

If the DFIG operates at a 10% power reserve, $\beta$ will increase by approximately 1.6°.

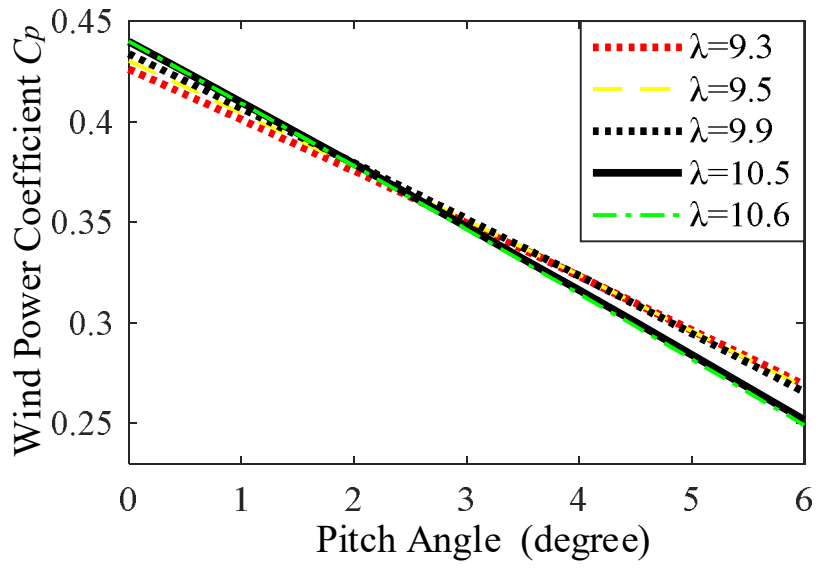

***Fig. 8** Wind power coefficient $C_p$ versus pitch angle under differ ent tip speed ratios.*

Further observing Fig. 7 in the low wind area, the $\Delta\beta$ is zero, where the de-loaded power is below 0.38$P_{nor}$; in the medium wind speed area, where the de-loaded power is between 0.38$P_{nor}$ and 0.9$P_{nor}$, the $\Delta\beta$ varies with de-loaded power. At high wind speeds, where the de-loaded power remains at 0.9 $P_{nor}$, $\Delta\beta$ remains at a constant value of 1.6°.

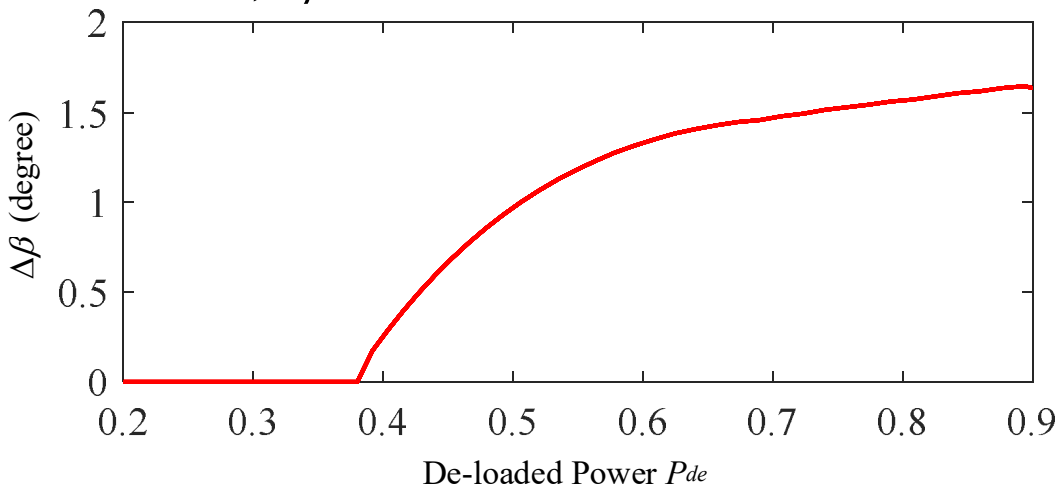

**Fig. 9** $\Delta\beta$ versus the de-loaded power curve

The $\Delta\beta$ versus the de-loaded power curve is described in Fig. 9. and the polynomial fitting function given in (15).

$$\Delta\beta=\begin{cases} 0 & P_{de} < 0.38 \\ 20.9P_{de}^3 - 48.17P_{de}^2 + 37.62P_{de} - 8.42 & 0.38 \le P_{de} < 0.9 \\ 1.6 & P_{de} = 0.9 \end{cases} \tag{15}$$

According to (15), $\Delta\beta$ can be obtained based on the de-loaded power $P_{de}$. Wind speed information is not required.

In the medium and high wind speed areas, the pitch angle must be adjusted to release more or less active power for participation in frequency regulation. The improved pitch control diagram is given in Fig. 10 below.

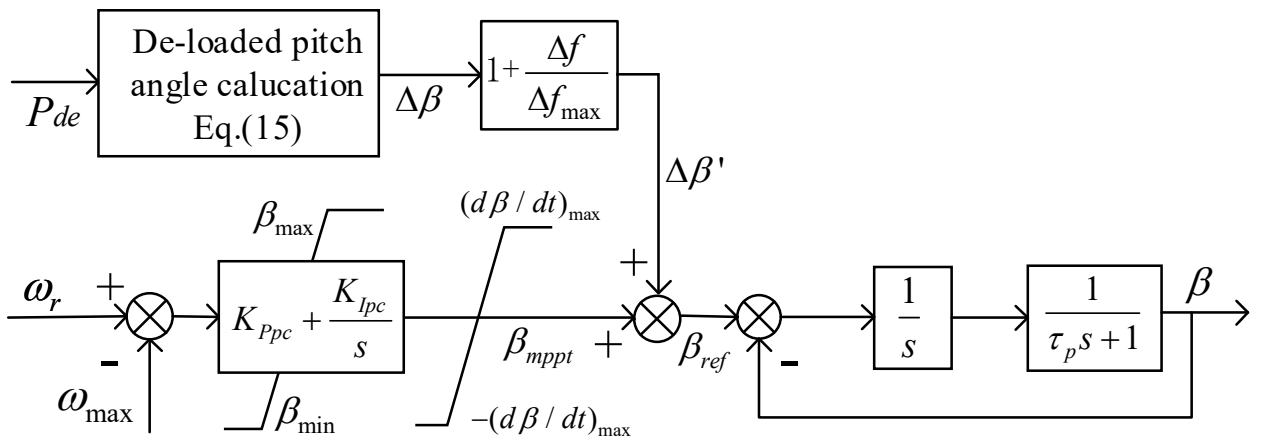

***Fig. 10** The proposed pitch angle control scheme*

As seen in Fig. 10, $\Delta\beta$ is obtained from equation (15), and a simple linear method is used to calculate the compensation pitch angle $\Delta\beta'$,

$$\Delta\beta'=\left(1+\frac{\Delta f}{\Delta f_{\max}}\right)\Delta\beta \tag{16}$$

$\Delta\beta'$ regulate with the system frequency. when $\Delta f$=$-\Delta f_{\max}$, $\Delta\beta'$ is equal to 0, and $\beta_{\text{ref}}=\beta_{mppt}$. When $\Delta f$=0, $\Delta\beta'$ is equal to $\Delta\beta$, and $\beta_{\text{ref}}$=$\beta_{de}$ (the pitch angle in de-loaded mode).

As shown in the modified pitch angle control description in Fig. 10, wind speed information is not required. The pitch angle can help the de-loaded operation and the frequency regulation of wind turbines in medium and high wind speed areas.

## 4. Case Study

In this section, simulations using MATLAB/ SIMULINK 2018 Student Suite Version, MathWorks, Natick, MA, USA are carried out to verify the proposed frequency regulation scheme's efficacy.

Fig. 11 shows the single bus model of the small isolated power system used in the paper. It includes static loads, one thermal plant, one hydropower plant, and one aggregated DFIG-based wind power plant. The total capacity of the power systems is 1250 MW.

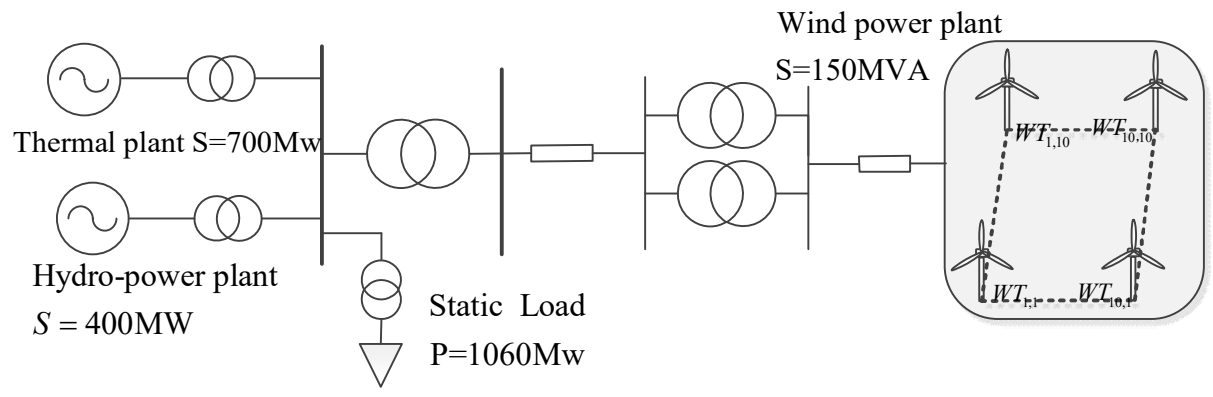

***Fig. 11** Single-line diagram of test power system for simulation*

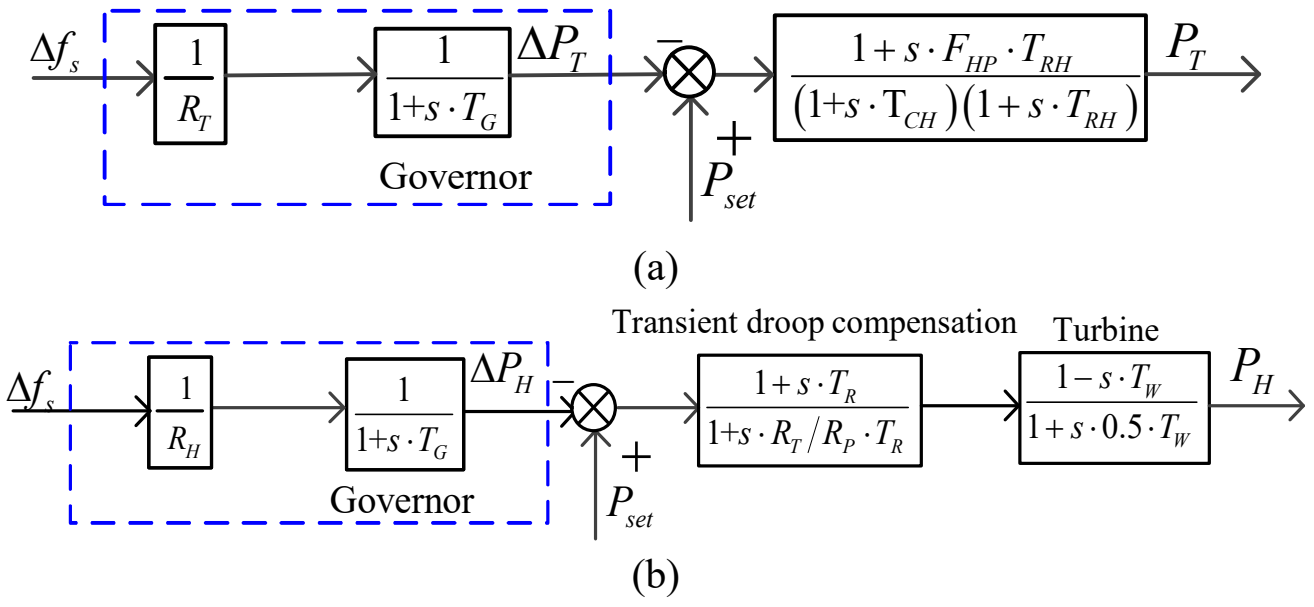

***Fig. 12** Governor-based models of conventional power plants*

Simplified governor-based models from [44] are used to simulate thermal and hydropower plants (see Fig. 12).

The droop characteristics of conventional plant speed governors have been enabled. The values of the most significant parameters are summarized in Appendix table2 and 3.

To simulate the wind power plant, an equivalent generator with 100 times the nominal power of one DFIG is assumed. The parameters of DFIG-based VSWTs are given in Appendix table.1 The performance of the proposed scheme for DFIG-based VSWTs is compared to that of MPPT, the conventional FRS with fixed gain under various wind speeds.

In traditional FRS with fixed gain, $K_w$ is set to 30 and 15, while $1/R_w$ is set to 24 and 7. Under medium wind speed conditions, when a larger gain is selected, the inertial control performance of the DFIG can be effectively improved while ensuring stable operation. In comparison, a smaller gain is selected to maximize the lowest frequency point (FN) while ensuring stable operation of all DFIGs under low wind conditions. It is worth noting that the values of large gain and small gain are just an example of traditional FRS. If the system changes, these values should be changed appropriately.

Cases 1, 2, and 3 refer to the constant wind speed in the low wind speed area, the medium wind speed area and the high wind speed area, respectively. In cases 4 and 5, the wind speed is assumed to be reduced at the instant of an event, from 9 to 7.5 m/s for 10 and 1 s, respectively. Case 6 is the random wind speed in low wind area. In all cases, if the rotor speed reaches $\omega_{min}$, the FRS (not including de-loaded control) are disabled by disconnecting the frequency measurement.

Compared with the decreased output power, DFIG-based VSWTs have more difficulty increasing the output power when the frequency dips. Thus, at 60 s, the system load suddenly increases by 0.1 pu and causes a frequency dip event for all cases.

### *4.1. Effects of Wind Speeds*

### *4.1.1 Case1: Low wind speed area*

Fig. 13 illustrates the results for a wind speed of 8 m/s in the low wind speed area, where the DFIG-based VSWTs only use overspeed control to realize de-loaded control. Fig. 13 shows the result of Case 1.

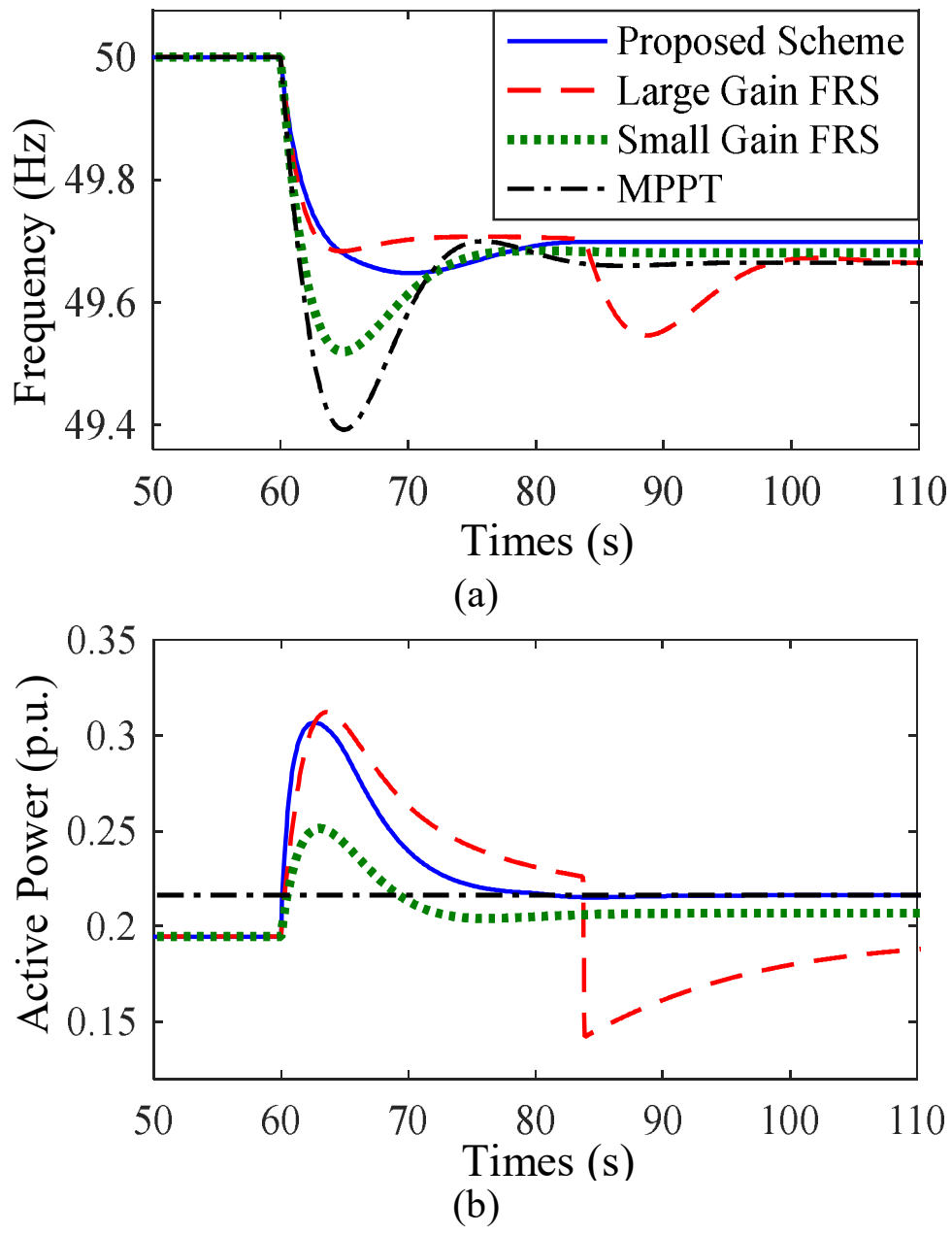

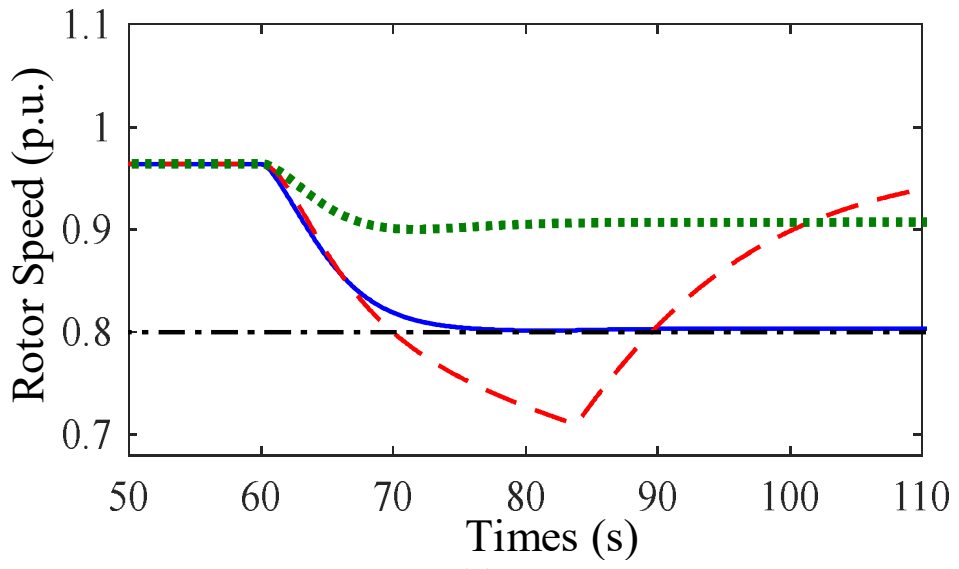

(c)

***Fig. 13*** *Results for Case 1. (a) System frequency; (b) Active power of DFIG; (c) Rotor speed of DFIG*

Before 60 s, for the traditional FRS and the proposed FRS, the initial power is maintained at $0.9P_m$, which is 0.194 pu. The output power of MPPT always remains at the maximum power of 0.216 pu.

For the large gain FRS, the system frequency reduction speed is slower than that of the small gain FRS and the proposed scheme. Because the large gain FRS releases more power in the initial stage of the grid, the rotor speed decreases more. As shown in Fig. 13(c), at 83.86 s, the speed of the DFIG reaches $\omega_{min}$. To protect the wind turbine, the frequency support control is disabled and causes sudden active power drops, which result in a secondary frequency dip. The first frequency nadir (FN) is 49.68 Hz, and the second FN is 49.55 Hz.

The small gain FRS can ensure that DFIGs operate in a safe rotor speed range. However, due to the limited support it provides, its peak power is only 0.25 pu, and FN is 49.46. The steady frequency is 49.68 Hz.

For the proposed scheme, the peak power is 0.31 pu, larger than the small gain and smaller than the large gain FRS. However, it can prevent the wind turbines from excessively releasing kinetic energy through the modified power versus rotor speed curves. The rotor speed almost reaches but does not exceed $\omega_{min.}$ The FN is 49.65 Hz. It is 0.13 Hz higher than that for the small gains. The steady frequency is 49.7 Hz, 0.02 Hz larger than the small gain FRS.

Fig. 14 illustrates the result of the operation at a wind speed of 10 m/s.

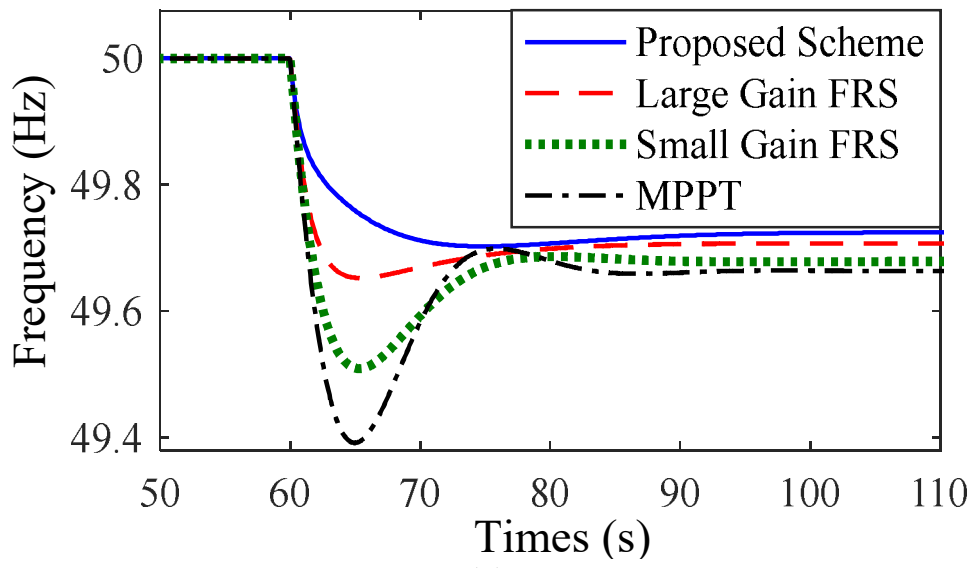

(a)

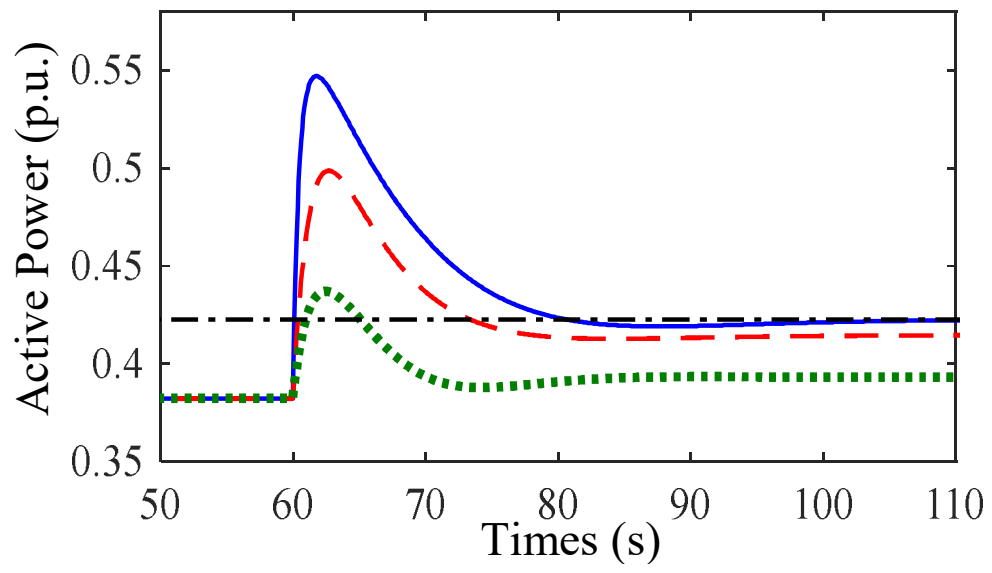

(b)

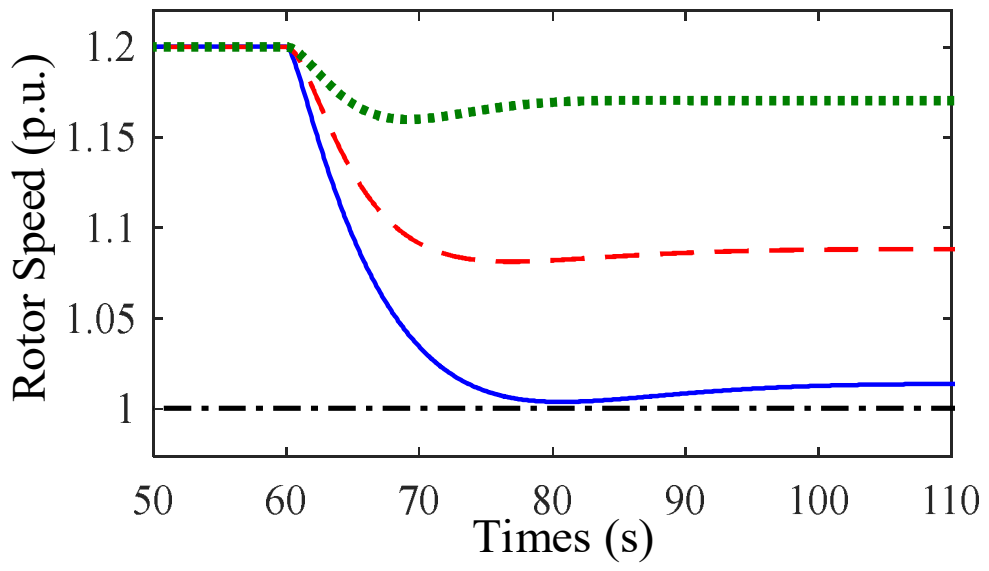

(c)

***Fig. 14*** *Results for Case 1. (a) System frequency; (b) Active power of DFIG; (c) Rotor speed of DFIG*

Due to the increase in wind speed, the traditional large-gain FRS's rotor speed does not exceed the lower limiter. Compared with the large gain FRS, the proposed scheme shows a better performance. The reason is that the proposed FRS has a potential self-adaption. With increasing wind speed, the active power for the frequency support increases.

The FN of the small gain FRS is 49.48 Hz, the steady frequency is 49.67 Hz, the FN of the large gain FRS is 49.65 Hz, the steady frequency is 49.7 Hz, and the proposed frequency is 49.7 Hz, which is 0.05 Hz larger than that of the large-gain FRS. The steady frequency is 49.72 Hz, which is 0.02 larger than the large gain FRS.

### *4.1.2 Case2: Medium wind speed area*

Fig. 15 illustrates the result of Case 2. In this case, the wind speed of the DFIG is 11 m/s in the medium wind area.

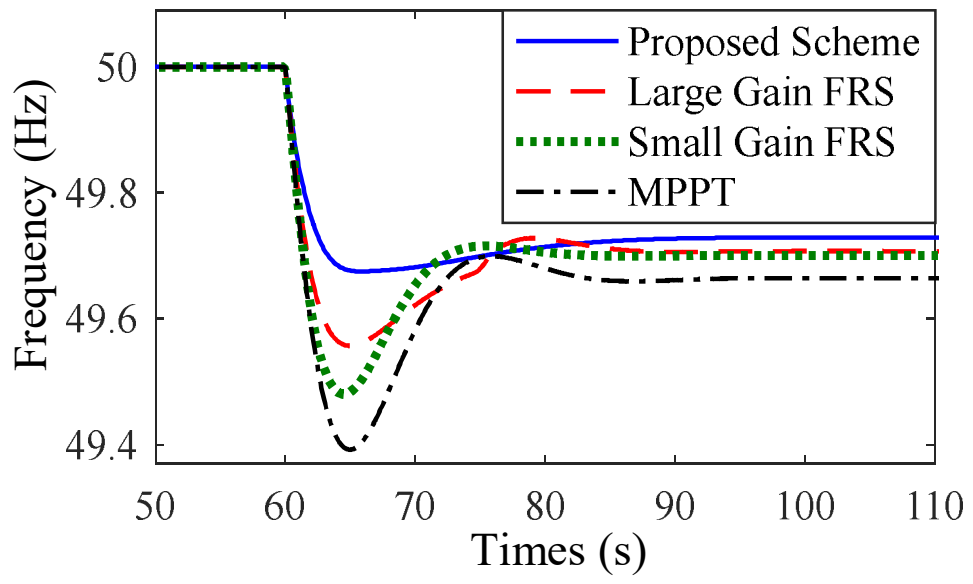

(a)

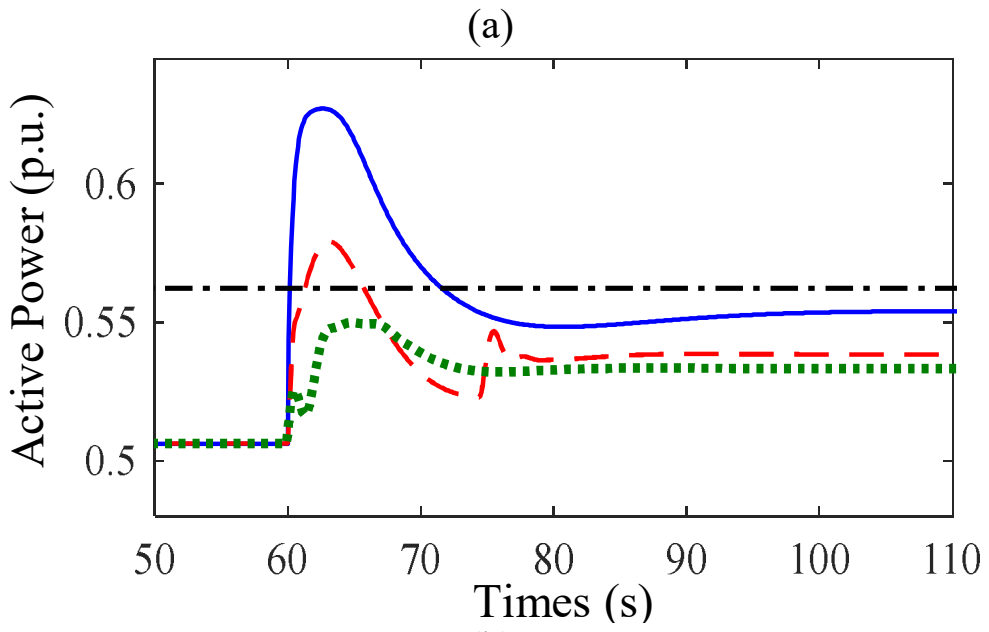

(b)

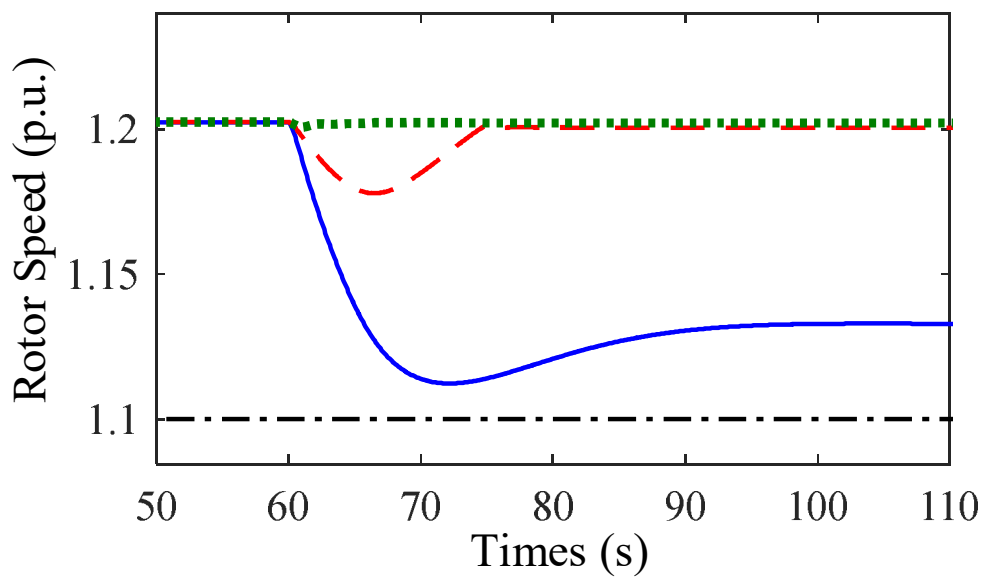

(c)

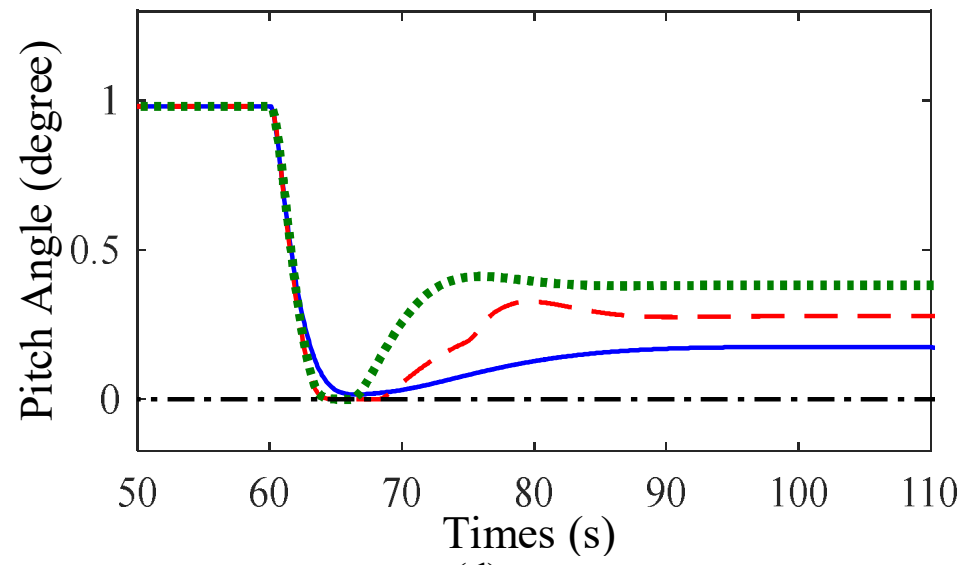

(d)

***Fig. 15*** *Results for Case 2. (a) System frequency; (b) Active power of DFIG; (c) Rotor speed of DFIG; (d) Pitch angle*

As shown in Fig. 15(a), the FN of the traditional small gain FRS is 49.48 Hz, which is 0.08 Hz higher than the FN of MPPT control but 0.08 Hz lower than the FN of the large gain FRS. The proposed scheme's FN is 49.73 Hz, which is 0.17 Hz higher than the large gain FRS. The proposed scheme's steady frequency is 49.7 Hz, 0.03 Hz larger than the large gain FRS and 0.04 Hz larger than the small gain FRS.

In Fig. 15(b), the output power of MPPT is always the maximum power of 0.5623 pu. Before 60 s, the traditional scheme and the proposed scheme maintain an initial power of 0.9. Initially, the proposed scheme's additional power is greater than that of the large gain and the small gain FRS. The designed inertia power curve is used to provide inertia support. The peak power of the small gain is 0.55 pu., that of the large FRS is 0.58 pu., and that of the proposed scheme is 0.627 pu.

In Fig. 15(c), the selected large- and small-gain FRSs can ensure that the DFIG works in the safe rotor speed range. The rotor speed of the proposed scheme converges to a lower value than the large gain FRS. Due to the potential self-adaption, the proposed scheme can adaptively release more KE from the DFIG.

As shown in Fig. 15(d), pitch angle control is activated to provide de-loaded and frequency support control. Before 60 s, the DFIG runs under de-loaded mode, and the pitch angle is approximately 0.98. After 60 s, the pitch angle decreases to release extra power for frequency support. The proposed scheme shows a deeper pitch drop than the others. It seems that the proposed scheme can better utilize rotor speed control to provide frequency support.

#### *4.1.3 Case3: High wind speed area*

Fig. 16 illustrates the result of Case 3. In this case, the wind speed is 13.6 m/s in the high wind speed area.

In high wind speed areas, the frequency regulation is depended on the pitch angle control. Fig. 16(c) shows that the rotor speed is near 1.21 pu for all the control schemes. In Fig. 16(d), the proposed scheme's pitch angle drops deeper than that of the traditional scheme. Therefore, the output power from the wind turbine is higher than that of the large gain FRS. The FN of the proposed scheme is better than that of the large gain FRS. As shown in Fig. 16 (a), the FN of the small gain FRS is 49.59 Hz, which is 0.03 Hz lower than the large FRS and 0.09 Hz lower than the proposed scheme.

Because of the different drop level of the pitch angle, the proposed scheme's steady frequency is the best, i.e., 49.79, while the steady frequency of the large-gain FRS is 49.77 Hz, and that of the small-gain FRS is 49.74.

The results of the three cases clearly show that the proposed scheme can provide better frequency support than the traditional scheme for the whole wind speed range, regardless of low, medium and high wind speeds. It has the potential to provide adaptive frequency support for different wind speeds and prevent overdeceleration of the DFIG.

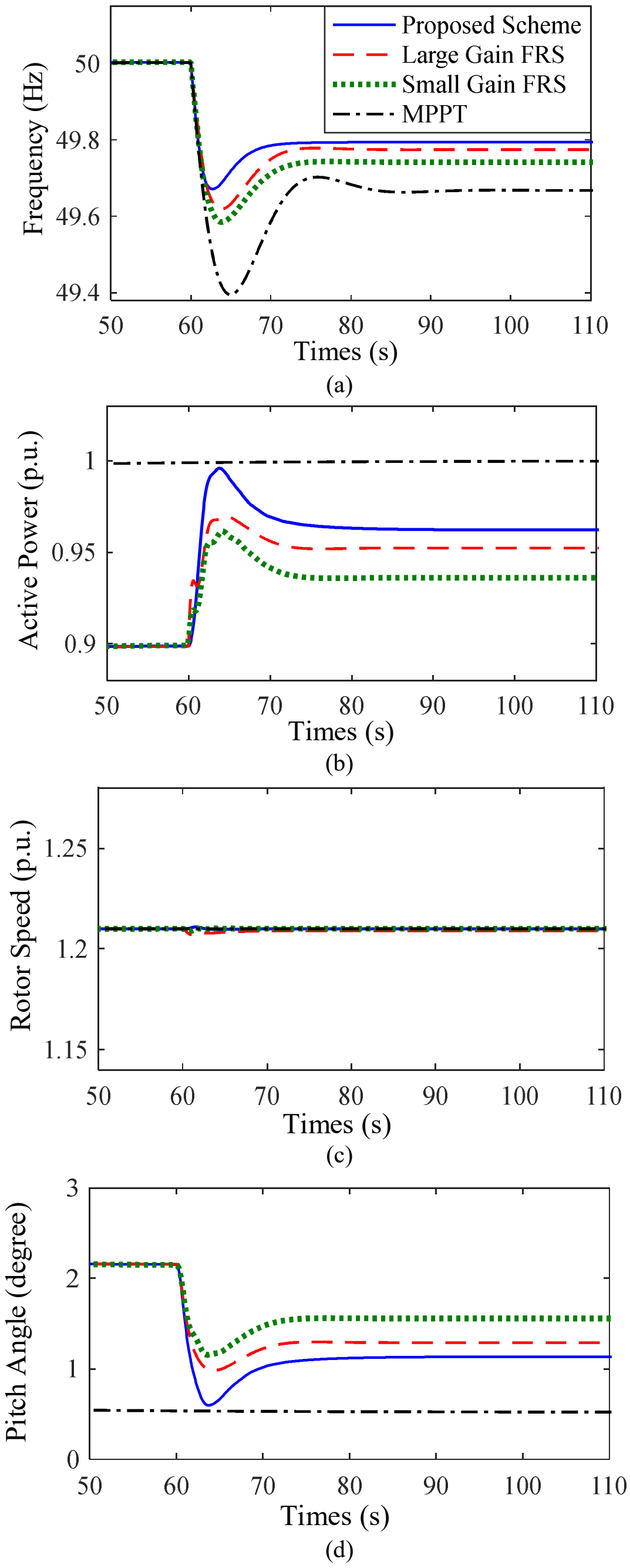

(d)

***Fig. 16*** *Results for Case 3. (a) System frequency; (b) Active power of DFIG; (c) Rotor speed of DFIG; (d) Pitch angle*

### *4.2. Effects of Varying Wind Speeds*

This section introduces the test results under variable wind speed conditions. The wind speed decreases at different time intervals, and the actual random wind speed is considered.

### *4.2.1 Case 4: Decreasing wind speed from 9 to 7.5 m/s for 10 s*

Fig. 17 illustrates the result of Case 4, where the wind speed decreases from 9 m/s to 7.5 m/s for 10 s at 60s.

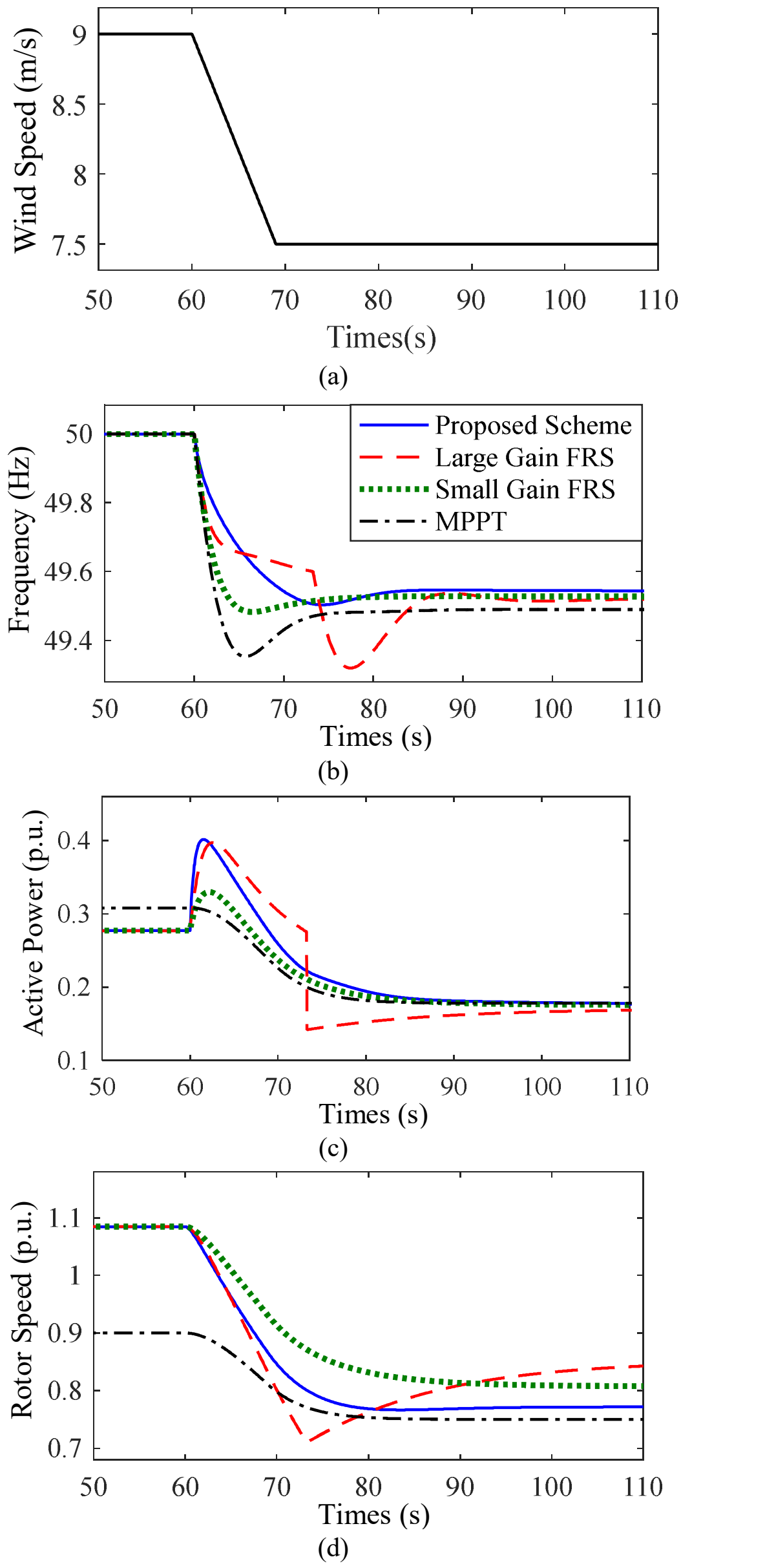

***Fig. 17*** *Results for Case4. (a) Wind speed; (b) System frequency; (c) Active power of DFIG; (d) Rotor speed of DFIG*

The proposed FRS prevents the rotor speed from exceeding $\omega_{min}$ and the rotor speed converges to 0.77 pu rapidly, which is a smaller value than the small gain FRS. Its FN is 0.02 Hz larger than the small gain and 0.18 Hz larger than the secondary frequency drop of the large gains FRS.

### *4.2.2 Case5: Decreasing Wind Speed from 9 to 7.5 m/s for 1 s*

Fig. 18 shows the result of Case 5. In this case, the wind speed was reduced from 9 to 7.5 m/s at 60 s for 1 s.

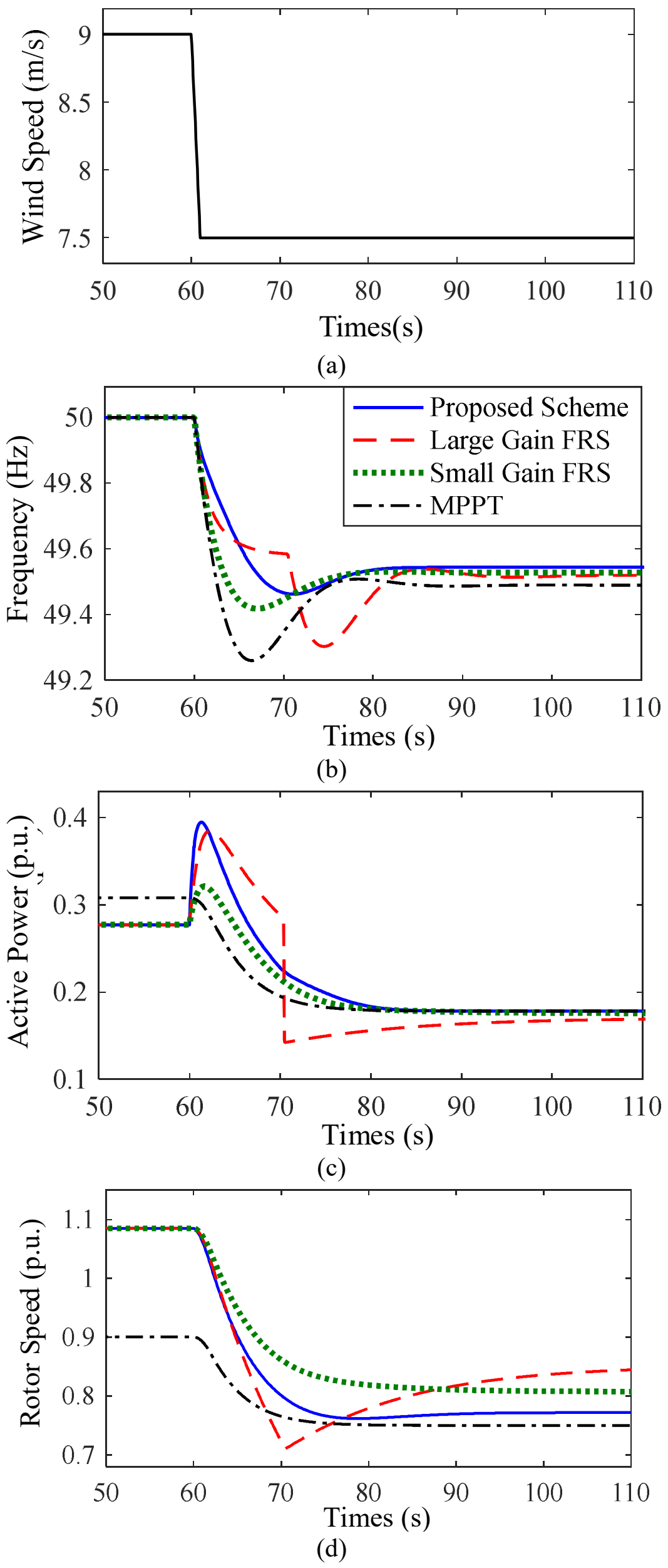

***Fig. 18*** *Results for Case 5. (a) Wind speed (b) System frequency; (c) Active power of DFIG; (d) Rotor speed of DFIG*

Compared with Case 4, the wind speed reduction is faster than that in Case 4. This results in a faster reduction in the rotor speed than Case 4. Along with the wind speed reduction, the output power of the wind turbine decreases.

For the small gain FRS, the wind turbine still has stable operation, and the FN is 49.41 Hz, 0.07 Hz lower than the small gain FRS in case 4.

However, for the large gain FRS, the rotor speed reaches $\omega_{min}$ at 70.6 s, which is 2.5 s earlier than case 4, and it causes a secondary frequency drop to occur. The secondary FN at 74.47s is 49.3 Hz.

Conversely, our proposed scheme can still successfully prevent overdeceleration even in this case by rapidly reducing the active power, as shown in Fig. 18(d). The speed converges to a safe value of 0.77 pu. The FN is 0.05 Hz higher than the small gain FRS. The steady-state value of the frequency is similar to the large gain.

### *4.2.3 Case6: Random wind speed*

Fig. 19 presents the simulation results of case 6. The actual measured wind speed (low wind speed area) in Northeast China. The load suddenly increases 0.1 pu at 60 s.

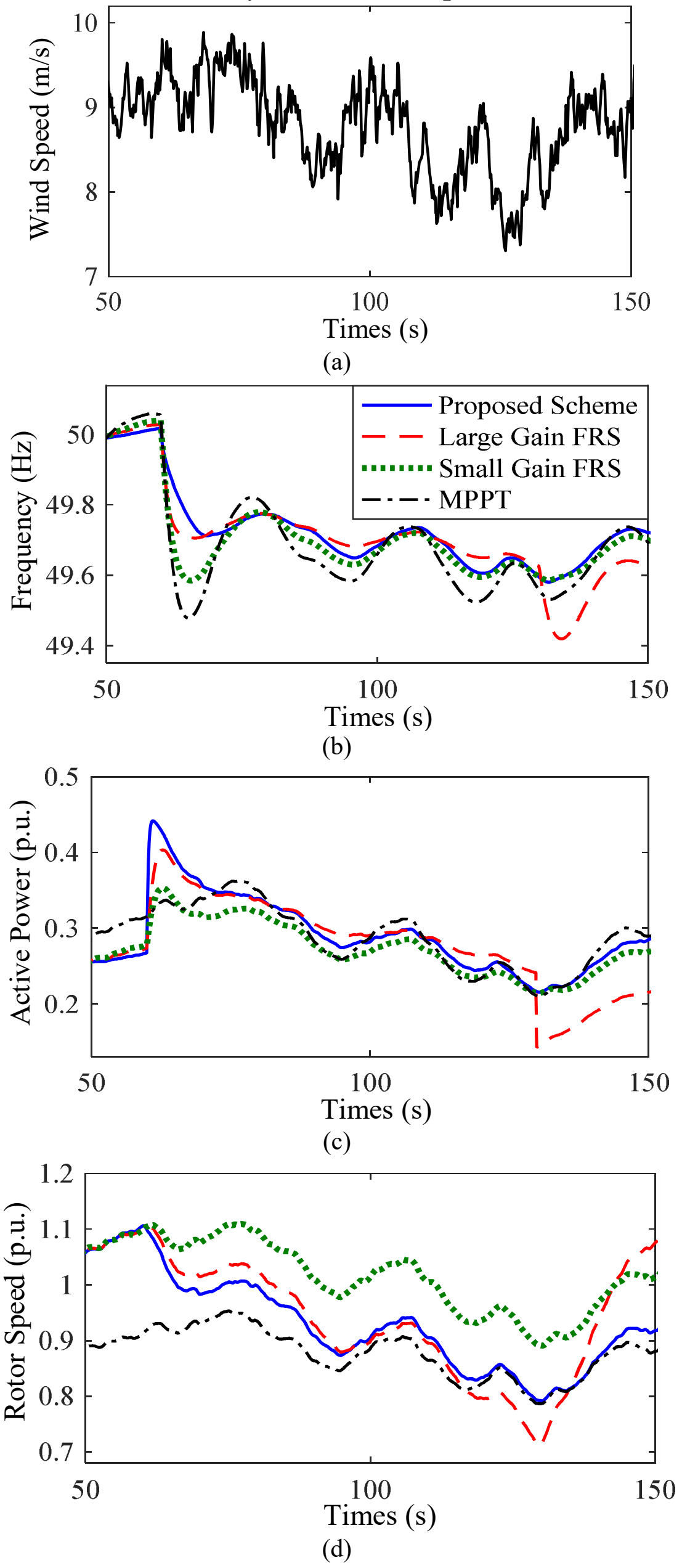

***Fig. 19*** *Results for Case 6. (a) Random wind speed; (b) System frequency; (c) Active power of DFIG; (d) Rotor speed of DFIG*

For MPPT, the active power is only affected by wind speed. The wind turbine does not participate in frequency regulation. The FN is 49.38 Hz.

The stable operation of the DFIG can be ensured in small gain FRS. However, the additional active power for the frequency regulation is insufficient. The FN is 49.59 Hz, which is 0.12 Hz lower than the proposed scheme.

Due to the excessive active power released in the large gain FRS, the rotor speed drops fast and exceeds $\omega_{min}$ at 130 s, causing a secondary frequency drop. The secondary FN is 49.42 Hz, which is the lowest value in this case.

From Fig. 19(b), before the FRC disabled, the large gain FRS shows a slightly better frequency response than the proposed FRS. However, it will result in an overdeceleration of the DFIG and cause a secondary frequency dip.

The simulation results of cases 4, 5, and 6 verify that even under variable wind speed conditions, the proposed scheme still prevents the wind turbine from over decelerating.

In summary, the pros and cons of traditional and proposed solution are listed in Table.1 below.

**Table 1** The pros and cons of the traditional solution and proposed solution

| | Traditional Scheme | Proposed Scheme |
|---|---|---|
| No control gains | auxiliary frequency controller is adopted and the parameter of auxiliary frequency controller need to compromise of the frequency regulation performance and the safe speed range the wind turbine operates. | The frequency response is generated for the modified power versus speed curve which does not contains any control gains |
| Potential self-adaptability | Fixed control gains: the gains of the virtual inertia and droop loops is fixed, and the inappropriate value of the gain will lead to the secondary frequency drop or insufficient power compensation.<br>Adaptive control gains: has adaptive control ability, But the calculation of the gains usually is complicated, which requires wind speed measurement or wind turbine parameters. | It has a potential self-adaptive frequency support ability. When the wind speed increases (or decrease), the output active power spontaneously increases (or decrease) |
| Not required wind speed information | The wind speed area decision is relied on the wind speed information.<br>The de-loaded pitch angle in medium and high wind area is required wind speed measurement or wind turbine parameters, and have a complicated calculation | Not required to collect wind speed information.<br>Fitting the off-line data about the pitch angle and the de-loaded power, the compensation value of the pitch angle changes adaptively with the change of frequency. |

## 5. Conclusion

With the high penetration of wind power generator, the frequency stability is still a critical issue in the power system because of the reduction of the inertia and frequency regulation capability. This paper proposes a novel frequency regulation scheme based on parameterized power curve for the de-loaded wind turbine during the whole wind speed range.

Compared with the existing scheme, the auxiliary frequency controller is replaced by a modified power versus rotor speed curve according to the frequency deviation and the rate of change of the frequency to provide both inertia and a droop frequency response. In the traditional scheme, the control gains of the auxiliary frequency controller are hard to select because of the compromising of the rotor speed safe and the frequency regulation performance. However, the proposed scheme does not contain any gains and can be applied to WTGs with different wind speeds. Further, it has a potential self-adaptability that the active power will automatically increase with wind speed increases.

An improved pitch angle control is designed which is generally adapts for the whole wind speed range without wind speed information. The curve of the de-loading pitch angle versus de-loaded power is fitting by off-line modeling. And then, the compensation pitch angle is simply calculated by frequency measurement information, avoiding the use of wind speed information and the complicated calculation.

The comparison simulation results under low wind speed, medium wind speed, high wind speed and variable wind speed were evaluated. The simulation results clearly show that the scheme has a better performance than the traditional frequency regulation strategies for improving the frequency response and preventing rotor speed over deceleration under variable wind conditions.

## Appendix

**Table 1** DFIG-based VSWTs parameters

| Parameter | Value |
|---|---|
| Nominal output power $P_{base}$ | 1.5 MW |
| Max./Min. torque of the generator $T_{g.max}/T_{g.min}$ | 1.07/0.05 |
| Based wind speed | 12m/s |
| Maximum power at based wind speed | 0.73 |
| Number of pole pairs $p$ | 2 |
| Air density $\rho$ | 1.255kg/m$^3$ |
| Radius of the rotor R | 38m |
| Nominal frequency | 50Hz |
| Min. /Max. blade pitch angle $\beta_{min}/\beta_{max}$ | 0° /45° |
| Maximum blade pitch angle rate $(d\beta/dt)_{max}$ | 2° /s |
| DFIG-PEC time constant $\tau_c$ | 20ms |
| Blade pitch servo time constant $\tau_p$ | 0s |
| Pitch controller gains $K_{Ppc}/K_{Ipc}$ | 500/0 |
| Speed controller gains $K_{Psc}/K_{Isc}$ | 0.3/8 |

**Table 2** Conventional generation units' parameters

| Parameter | Value |
|---|---|
| Thermal generator capacity $P_t$ | 700 MW |
| Speed droop $R_T$ | 0.05 |
| Governor time constant $T_G$ | 0.2s |
| Fraction of power generated $F_{HP}$ | 0.3p.u. |
| Reheat time constant $T_{RH}$ | 7s |
| Turbine time constant $T_{CH}$ | 0.3s |
| Inertia constant time $H_T$ | 5s |

**Table 3** Hydro-power plant parameters

| Parameter | Value |
|---|---|
| Thermal generator capacity $P_H$ | 400 MW |
| Governor time constant $T_G$ | 0.2s |
| Speed droop $R_H$ | 0.05 |
| Reset time $T_R$ | 5s |
| Temporary droop $R_T$ | 0.38 |
| Permanent droop $R_P$ | 0.05 |
| Water starting time | 1s |
| Inertia constant $H_H$ | 3s |

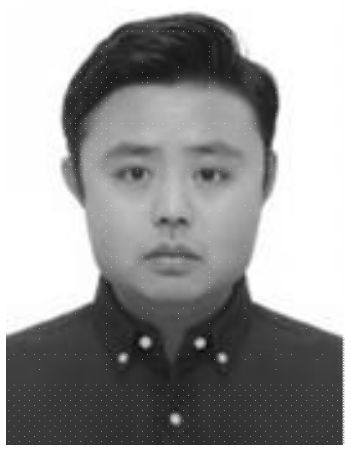

**CHENG ZHONG** (M'20) was born in JiangXi, China, in 1985. He received the B.S. degree from the Harbin Institute of Technology,Harbin,China, in 2007 and the M.S. and Ph.D. degrees in power systems and automation from China Agricultural University, Beijing, China, in 2010 and 2014, respectively. From January 2019 to February 2020, he held an academic visitor at the University of Birmingham, Birmingham, UK.

He is currently an Assistant Professor with the School of Electrical and Electronic Engineering, Northeast Electric Power University, JiLin, China, where he has been since 2014. His current research interests include the area of power system distribute intelligent control, power system frequency regulation by renewable power.

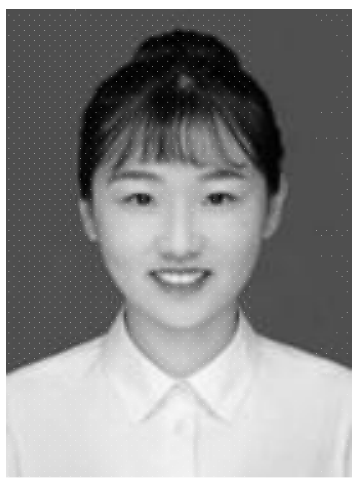

**YUEMING LV**(S'20) was born in JiLin, China, in 1997. She received the B.S. degree from Northeast Electric Power University, in 2019. She is currently pursuing the M.E. degree in electrical engineering from Northeast Electric Power University. Her research interest includes power system frequency regulation.

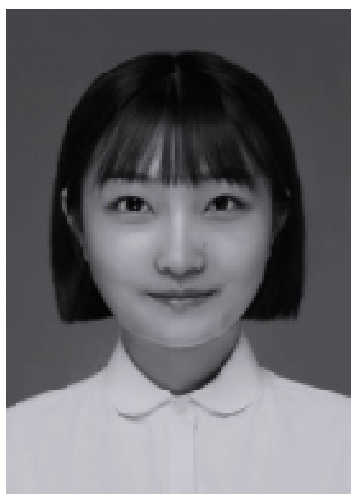

**HUAYi LI** (S'20) was born in Baoding, China, in 1998. She received the B.S. degree from Normal China Electric Power University Science and Technology College, in 2019. She is currently pursuing the M.E. degree in electrical engineering from Northeast Electric Power University. Her research interest includes active frequency control of high permeability photovoltaic island microgrid.

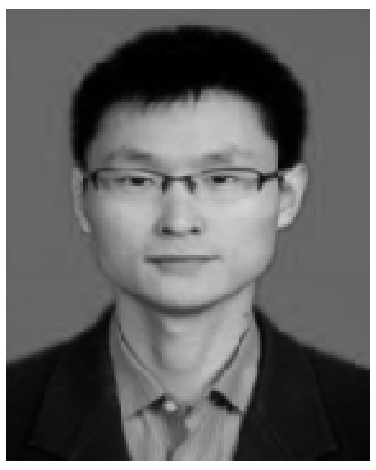

**JIKAI CHEN** (Member, IEEE) received the Ph.D. degree in electrical engineering from the Harbin Institute of Technology, Harbin, China, in 2011.He is currently an Associate Professor with the School of Electrical Engineering in Northeast Electric Power University. His current research interests include HVdc control technologies, analysis and control of power quality, and modeling analysis and control of power electronics dominated power systems.

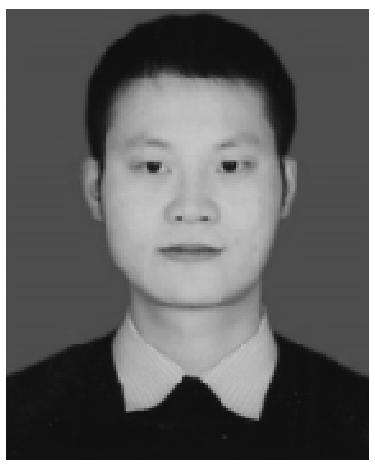

**YANG LI** (S'13 – M'14 – SM'18) was born in Nanyang, China. He received the Ph.D. degree in electrical engineering from North China Electric Power University (NCEPU), Beijing, China, in 2014.

He is currently an Associate Professor with the School of Electrical Engineering, Northeast Electric Power University, Jilin, China. From January 2017 to February 2019, he held a China Scholarship Council (CSC)-funded postdoctoral postition at the Argonne National Laboratory, Lemont, USA. His research interests include power system stability and control, renewable energy integration, and smart grids. He also serves as an Associate Editor for IEEE ACCESS.